\documentclass{svmult}
\usepackage{graphicx} 
\usepackage{url}
\newcommand{\beq}{\begin{equation}}
\newcommand{\eeq}{\end{equation}}

\begin{document}

\title*{Multiscale Methods}
\author{Enn Saar\inst{1,2}}
\institute{Tartu Observatoorium, T\~oravere, 61602 Tartumaa, Estonia, \\
\url{saar@aai.ee}\\
\and
Observatori Astron\`omic, Universitat de Val\`encia, 
Apartat de Correus 22085, E-46071 Val\`encia, Spain, \\
\url{Enn.Saar@uv.es}}

\maketitle

Bernard Jones told you about multiresolution analysis in his wavelet
lectures. This is a pretty
well formalized and self-contained area of wavelet analysis. Multiscale
methods form a wider and less well defined area of tools and approaches;
the term is used, e.g., in numerical analysis, but the main range of
multiscale methods are based on application of wavelets.

In this this lecture I shall explain how to carry out multiscale 
morphological analysis of cosmological density fields. For that,
we have to use  wavelets to decompose the data into different frequency bands, 
to calculate densities, and to describe morphology.

Let us start with wavelet transforms.

\section{Wavelet Transforms}

The most popular wavelet transforms are the orthogonal fast transforms,
described by Bernard Jones. For morphological analysis, we need
different transforms. The easiest way to understand wavelets is to
start with continuous wavelet transforms. 

\subsection{Continuous Wavelet Transform\label{sec:cont}}

\index{wavelet transform!continuous}
The basics of wavelets are most easily understood in the case of
one-dimensional signals (time series or data along a line). 
The most commonly used decomposition of such a signal (f(x)) into
contributions from different scales is the Fourier decomposition:
\[
	\hat{f}(k)=\int_{-\infty}^{\infty}f(x)\exp(-\I\,kx)\,dx\;.
\]
The Fourier amplitudes $f(k)$ describe the frequency content of a signal.
They are not very intuitive, however, as they depend on the behaviour of
a signal as a whole; e.g., if the signal is a density distribution along
a line, then all the regions of the universe where this line passes through,
contribute to it.  Fourier modes are homeless.

	For analyzing the texture of images and fields, both scales and
positions are important. The right tools for that are continuous wavelets.
A wavelet transform of our signal $f(x)$ is
\[
W(a,b)=\frac1{\sqrt{a}}\int_{-\infty}^\infty 
	f(x)\psi\left(\frac{x-b}{a}\right)\,dx\;,
\]
where $\psi(x;a,b)$ is a \index{wavelet!profile}
wavelet profile. Here the argument
$b$ ties the wavelet to a particular location, and $a$ is the scale factor.

\begin{figure}
\centering
\resizebox{0.4\textwidth}{!}{\includegraphics*{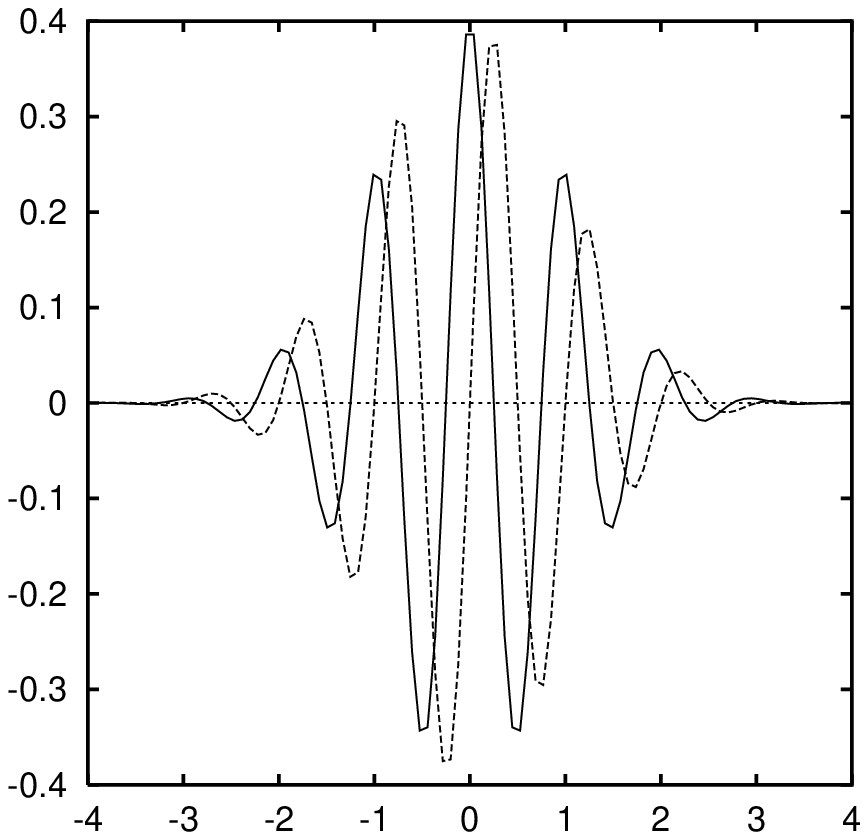}}
\resizebox{0.4\textwidth}{!}{\includegraphics*{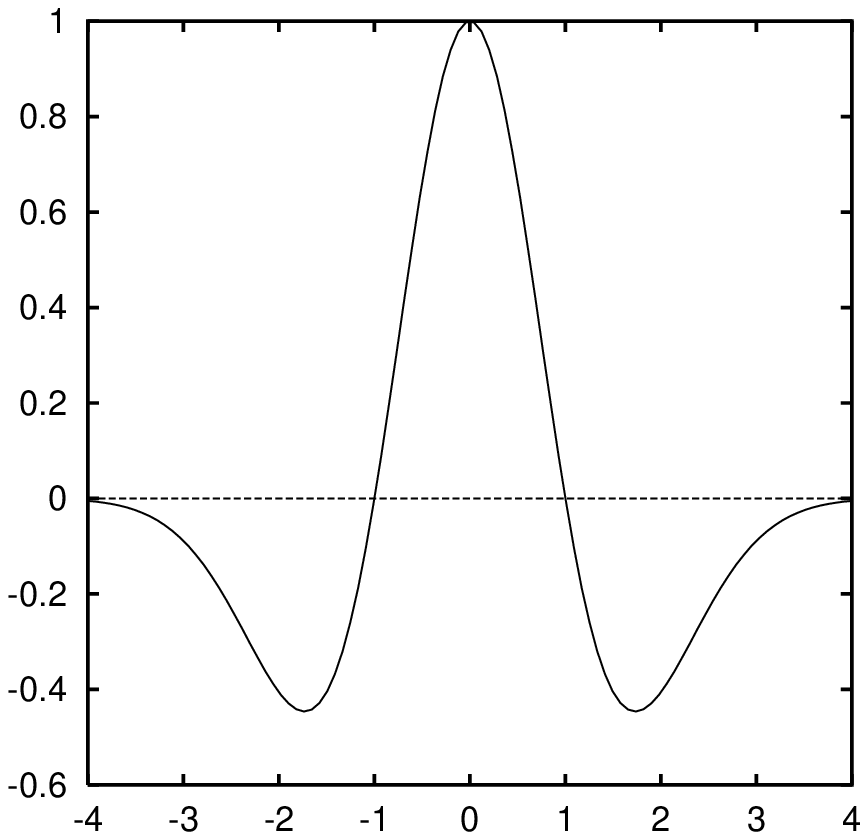}}\\
\caption{Left -- Morlet wavelet (solid line shows its real part,
dashed line -- the imaginary part), right -- (a real-valued ) Mexican hat 
wavelet
\label{fig:example_wavelets}}
\end{figure}

In order to be interesting (and different from the Fourier modes),
typical wavelet profiles have a compact support. Two popular wavelets are
shown in Fig.~\ref{fig:example_wavelets} -- the Morlet wavelet, and the Mexican
hat wavelet (see the formulae in Bernard Jones lecture). The Morlet
wavelet is a complex wavelet.

\begin{figure}
\centering
\resizebox{0.7\textwidth}{!}{\includegraphics*{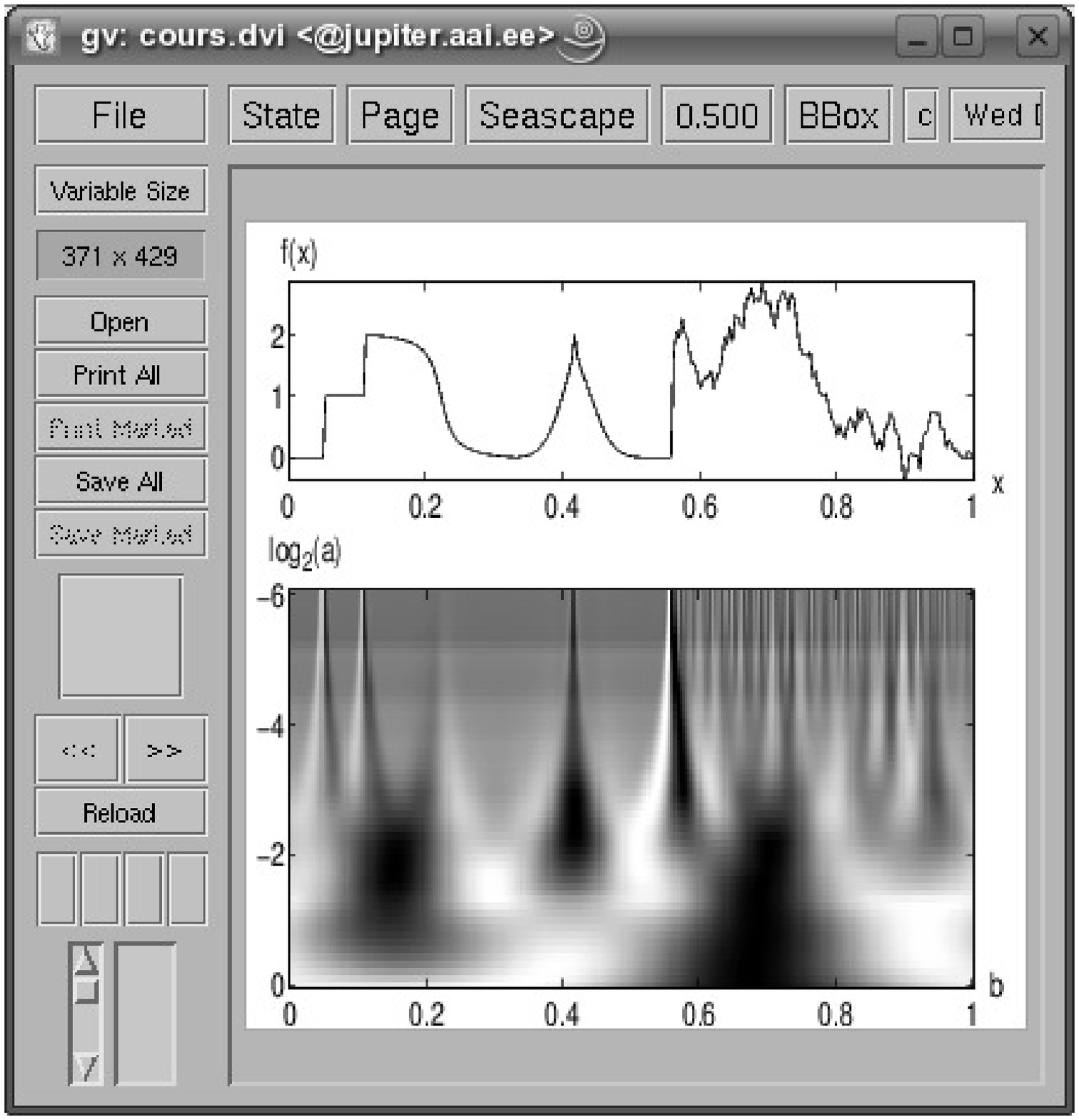}}
\caption{Example (continuous mexican hat wavelet)\label{fig:mexhat}}
\end{figure}

Continuous wavelets are good for finding sharp edges and singularities of
functions. An example of that is given in Fig.~\ref{fig:mexhat}. The
brightness-coded wavelet amplitudes (mexican-hat wavelet) for the
upper-panel skyline show features at all scales.

	This example shows us also the information explosion inherent to
continuous wavelets -- a function $f(x)$ gives rise to a two-argument
wavelet amplitude $W(a,b)$; the collection of continuous wavelet amplitudes 
is heavily redundant. If the wavelet is well-behaved (\ref{waveint}),
the wavelet amplitudes can be used to restore the original function:
\beq
\label{restor}
f(x)=\frac1{C_\psi}\int_0^\infty \int_{-\infty}^\infty 
\frac1{\sqrt{a}}W(a,b)\psi^\star\left(\frac{x-b}{a}\right)
\frac{da\,db}{a^2}\;,
\eeq
\pagebreak

\noindent where
\[
C_\psi=\int_0^\infty \frac{\hat{\psi}^\star(\nu)\hat{\psi}(\nu)}{\nu}d\nu\;
\]
($\psi^\star$ is a complex conjugate of $\psi$, and $\hat{\psi}$ is the
Fourier transform of $\psi$).
This constant exists ($C_\psi<\infty$), when
\beq
\label{waveint}
\hat{\psi}(0)=0\Rightarrow
\int_{-\infty}^\infty \psi(x)\,dx=0\;.
\eeq
This is the only requirement that a continuous-transform wavelet 
has to satisfy. Equation (\ref{restor}) shows that we need to know the
wavelet amplitudes of all scales to reconstruct the original function.
We also see that large-scale wavelet amplitudes are heavily downweighted
in this reconstruction.

\subsection{Dyadic Wavelet Transform}

\index{wavelet transform!dyadic}
It is clear that the information explosion inherent in application of 
continuous wavelets has to be restrained.
The obvious way to proceed is to consider computational
restrictions.

First, computations need to be carried out on a discrete coordinate
grid $b_i$. Second, if we look at Fig.~\ref{fig:mexhat} we see
that the wavelet amplitudes change, in general, smoothly with scale.
This suggests that a discrete grid of scales could suffice to
analyze the data. While the obvious choice for the coordinate grid
is uniform, the scale grid is usually chosen logarithmic, in hope to
better catch the scale-dependent behaviour. This
discretisation generates much less data than the continuous transform
did, only $N\times J$, where $N$ is the size of the original data
(the number of grid points), and $J$ is the number of different
scales. The most popular choice
of the scale grid is where neighbouring scales differ by a
factor of two. Other choices are possible, of course, but this
choice is useful in several respects, as we shall see below.

Such a wavelet transform is called dyadic. As we saw above, any
compact function with the zero mean could be used as an analyzing
wavelet. \index{wavelet!analyzing} 
In the case of a discrete wavelet transform, we, however,
have a problem -- can we restore the original signal, given all the
wavelet amplitudes obtained? It is not clear at first sight, as the
restoration integral for a continuous wavelet (\ref{restor}) contains
contributions of all wavelet scales.

The answer to that is yes, given that the frequency axis is completely
covered by the wavelets of dyadic scales (the wavelets form a so-called
frame). This requirement is referred to as the perfect reconstruction condition;
I shall show a specific example below.

Let us rewrite now the wavelet transform, for finite grids.
As we have heard about the multiresolution analysis already, we shall
try to introduce both the smooth and the detail part of the transform.
Let the initial data be $a_0(m)$ (the index $0$ shows the transform
order, and the argument $m$ -- the grid point). The smoothing operation and
the wavelet transform proper are convolutions, so they can be written
 \beq
 \label{dyadwave1}
a_{j+1}(m)=\sum_k h_j(k)a_j(m+k)\;, \qquad d_{j+1}(m)=\sum_k g_j(k)a_j(m+k)\;,
 \eeq
where the filters $h_j(k)$ and $g_j(k)$ are different from zero for a few
values of $k$ around $0$ only (the wavelet transform is 
local). Applying recursively this rule, we can find wavelet transforms of
all orders, starting from the data ($a_0(\cdot)$).

Now, we should like to be able to restore the signal, knowing the
$a_j(\cdot)$ and $d_j(\cdot)$. As the previous filters were linear,
restoration should also be linear, and we demand:
 \beq
 \label{dyadrecon}
a_j(m)=\sum_k \tilde{h}_j(k)a_{j+1}(m+k) + \sum_k \tilde{g}_j(k)d_{j+1}(m+k)\;.
 \eeq
It can be proved that these rules will work for any dyadic wavelet
that satisfies the so-called unit gain condition, if the filters
are upsampled (see next section).

If you recall (or look up) Bernard Jones' lecture, you will notice that the
rules (\ref{dyadwave1},\ref{dyadrecon}) look the same as the rules for
bi-orthogonal wavelets. The only difference is that for bi-orthogonal wavelets
the filters have to satisfy an extra, so-called aliasing cancellation condition,
arising because of the downsampling of the grid.

\subsection{\`A Trous Transform}

\index{wavelet transform!{\`a trous}}
This downsampling leads us to the next problem. The change of the frequency
(scale) between wavelet orders when bi-orthogonal or orthogonal wavelet
transforms are used, is achieved by \index{downsampling}
downsampling -- choosing every other data
point. Applying the same wavelet filter on the downsampled data set is
equivalent to using the twice as wide filter on the original data. Now, in
our case the grid is not diluted, and all points participate for all wavelet orders. The obvious solution here is to upsample the filter -- to introduce
zeros between the filter points. This doubles the filter width, and it is also
useful for the computational point of view, as the operations count for
the convolutions does not depend now on the wavelet order. Because of these
zeros (holes), such a dyadic transform is called ``\emph{\`a trous}''
('with holes' in French; the transform was introduced by French
mathematicians, Holschneider et al. (\cite{ESholtz})).

In Fourier language, such an upsampling 
\index{upsampling} halves the frequency:
\[
\hat{h}_1(\omega)=\hat{h}_0(2\omega)\;,\qquad 
\hat{h}_j(\omega)=\hat{h}(2^j\omega)\;,
\]
where $\omega$ stands for the frequency,
and $h(k)\equiv h_0(k), \hat{h}(\omega)\equiv \hat{h}_0(\omega).$
Although we describe localized transforms, it is useful to use Fourier 
transforms, occasionally  -- convolutions are frequent in wavelet transforms,
and convolutions are converted to simple multiplications in Fourier space. 
Now, in coordinate space, the explicit expression for the 
upsampled filter $h(\cdot)$ is:
 \[
h_j(k')=h(k)\delta(k'-k\cdot2^j)\;,
 \]
saying simply that the only non-zero components of the smoothing
filter for the order $j$ are those that have the index $k\cdot2^j$,
and they are determined by the original filter values $h(k)\equiv
h_0(k)$. Let us write now the \index{smoothing sum}
smoothing sum again:
 \beq
 \label{dyadsm}
a_{j+1}(m)=\sum_{k'}h_j(k') a_j(m+k')=\sum_k h(k) a_j(m+2^jk)\;,
 \eeq
where we retained only the non-zero terms in the last sum. The last
equality is of the form usually used in \emph{\`a trous} transforms;
the holes are returned to the data space again. However, the
procedure is different from the bi-orthogonal transform, as we find
the scaling and wavelet amplitudes for every grid point $m$, not for
the downsampled sets only. The wavelet rule with holes reads
 \beq
 \label{dyadw}
d_{j+1}(m)=\sum_k g(k) a_j(m+2^jk)\;,
 \eeq
where, obviously, $g(k)\equiv g_0(k)$.

Let us now construct a particular \emph{\`a trous} transform, starting
backwards, from the reconstruction rule (\ref{dyadrecon}). In
multiresolution language, this rule tells us that the
approximation (sub)space $V_j$ where the ``smooth functions'' live,
is a direct sum of two orthogonal subspaces of the next order
(the formula is for projections, but it follows from this fact).
Let us take it in a much simpler way, literally, and demand:
 \beq
 \label{atrecon}
a_j(m)=a_{j+1}(m)+d_{j+1}(m)\;,
 \eeq
or $\tilde{h}(k)=\tilde{g}(k)=\delta_{0k}\;$.
This is a very good choice, as applying it recursively we get
 \beq
 \label{atsum}
a_0(m)=a_J(m)+\sum_{j=J}^{j=1} d_j(m)\;,
 \eeq
meaning that the data is decomposed into a simple sum of contributions
of different details \index{wavelet transform!order}
(wavelet orders) and the most smooth picture.
As there are no extra weights, these detail spaces have a direct
physical meaning, representing the life in the full data space at
a given resolution.

The condition (\ref{atrecon}) gives us at
once the formula for the wavelet transform
 \beq
 \label{atwave}
d_{j+1}(m)=a_j(m)-a_{j+1}(m)=\sum_k\left[\delta_{0k}-h(k)\right]a_j(m+k)\;,
 \eeq or $g(k)=\delta_{0k}-h(k)\;$.

So far, so good. Now we have only to choose the filter $h(k)$ to
specify the transform. As the filter is defined by the scaling
function $\phi(x)$ via the \index{two-scale equation}
two-scale equation
 \beq
 \label{twoscale}
\phi(x/2)=2\sum_k h(k)\phi(x-k)\;,
 \eeq
 meaning simply that the
scaling function of the next order (note that $f(x/2)$ is only half
as fast as $f(x)$) has to obey the smoothing rule in (\ref{dyadsm}) exactly
(the space where it lives is a subspace of the lower order space).
Different normalizations are used; the coefficient 2 appears here if
we omit extra coefficients for the convolution (\ref{twoscale}) in
the Fourier space:
 \[
\hat{\phi}(2\omega)=\hat{h}(\omega)\hat{\phi}(\omega)
 \]
(recall that the coordinate space counterpart of $\hat{f}(2\omega)$
is $f(x/2)/2$).  Obviously, not all
functions satisfy the two-scale equation (\ref{twoscale}), but a useful
class of functions that do are box splines. 

\subsection{Box Splines}

\index{box spline}
Box splines are easy to obtain -- an $n$-th degree box spline
$B_n(x)$ is the result of $n+1$ convolutions of the box profile
$B_0(x)=1, \quad x\in[0,1]$ with itself. Some authors like to shift
it, some not; we adopt the condition that the convolution result is
centred at 0 when $n$ is odd, and at $x=1/2$ when $n$ is even. This
convention gives a simple expression for the Fourier transform of
the spline:
 \beq
 \label{fourspl}
\hat{B_n}(\omega)=\left(\frac{\sin(\omega/2)}{\omega/2}\right)
  \exp\left(-\I\,\varepsilon\omega/2\right)\;,
 \eeq
where $\varepsilon=1$, if $n$ is even, and 0 otherwise. You can easily derive
the formula yourselves, recalling that the Fourier transform of a $[-1,1]$ box
is the sinc($\cdot$) function, and using the rule for the argument shifts.

Box splines have, just to start, several very useful properties. First,
they are compact; in fact, they are the most compact polynomials for a
given degree ($B_n(x)$ is a polynomial of degree $n$). Second, they are
\index{box spline!interpolating}
interpolating,
 \beq
\label{interpol}
\sum_k B_j(x-k)=1\;,
 \eeq
a necessary condition for a scaling function. And box splines satisfy
the two-scale equation (\ref{twoscale}), with 
\beq
\label{hk}
h(k)=2^{-(n+1)}{n\choose k}\;,
\eeq
where $n$ is the degree of the spline (see de Boor (\cite{ESdeboor}). 
This formula is written for 
unshifted box splines, and here the index $k$ ranges from 0 to $n$.
It is easy to modify (\ref{hk}) for centred splines; e.g., for
centred box-splines of an odd degree $n$  the index $k$ ranges from
$-(n+1)/2$ to $(n+1)/2$, and we have to replace $k$ at
the right-hand side of (\ref{hk}) by $k+(n+1)2$.

\index{box spline!{$B_3$}}
I do not intend to be original, and shall choose the $B_3$ box
spline for the scaling function. This is the most beloved box spline
in astronomical community; see, e.g., the monograph by Jean-Luc
Starck and Fion Murtagh (\cite{ESstarck06}) for many examples. As any
spline, this can be given by different polynomials in different
coordinate intervals; fortunately, a compact expression exists:
 \beq
 \label{b3}
 \phi(x)=B_3(x)=
	\frac{1}{12}\left(|x-2|^3-4|x-1|^3+6|x|^3-4|x+1|^3+|x+2|^3\right)\;.
 \eeq
This function is identically zero outside the interval $[-2,2]$. 
Formula (\ref{hk}) gives us the filter $h(k)$:
 \beq
 \label{b3h}
 h(k)=(1/16,1/4,3/8,1/4,1/16)\;, \qquad k\in[-2,2]\;.
 \eeq
In order to obtain the associated wavelet $\psi(\cdot)$, we have to return to
our recipe for calculating the wavelet coefficients (\ref{atwave}).
These coefficients are, in principle, convolutions 
(multiplications in Fourier space):
 \beq
\hat{d}_{i+1}(\omega)=\hat{\psi}(\omega)\hat{a}_i(\omega)\;,
 \eeq
\pagebreak

\noindent So, (\ref{atwave}) gives us
 \[
\hat{\psi}(\omega)\hat{a}_i(\omega)=\hat{\phi}(\omega/2)\hat{a}_i(\omega)-
	\hat{\phi}(\omega)\hat{a}_i(\omega)\;,
 \]
or,
 \beq
 \label{psifour}
\hat{\psi}(\omega)=\hat{\phi}(\omega/2)-\hat{\phi}(\omega)\;.
 \eeq
For coordinate space, the above expression transforms to
 \beq
 \label{psi}
 \psi(x)=2\phi(2x)-\phi(x)\;.
 \eeq

\begin{figure}
\centering
\resizebox{0.6\textwidth}{!}{\includegraphics*{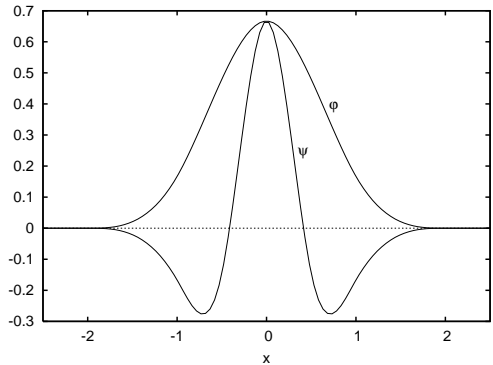}}
\caption{The $B_3$ scaling function ($\phi$) and its
associated wavelet ($\psi$)\label{fig:b3}}
\end{figure}

 Both the $B_3$ box spline (the scaling function $\phi(x$) 
and its associated
 wavelet $\phi(x)$ are shown in Fig.~\ref{fig:b3}. 
 
I cannot refrain from noting that the last exercise was,
strictly speaking, unnecessary. We could have proceeded with our wavelet
transform after obtaining the filter $h(k)$ (\ref{b3h}). But is nice
to know the wavelet by face and to get a feeling what our algorithms really
do.

Now I can also show you the Fourier transform of the wavelet (\ref{psifour}), to
reassure you that such a wavelet can be built and that it does not
leave gaps in the frequency axis. 
We see, first, that the filter peaks
at $\omega\approx\pi$, giving for its characteristic wavelength
$\lambda=2\pi/\omega=2$ (grid units). Second, we see that
neighbouring wavelet orders overlap in frequency. \index{frequency overlap}
 The reason for that is that our wavelets are not orthogonal. So
we loose a bit in frequency separation, but gain in spatial
resolution.

\begin{figure}
\centering
\resizebox{0.6\textwidth}{!}{\includegraphics*{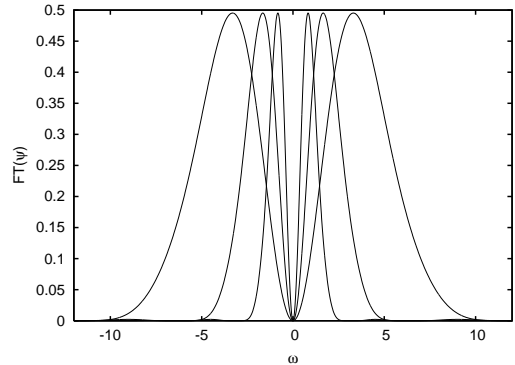}}
\caption{The Fourier transforms of the three subsequent orders of the 
$B_3(\cdot)$-associated wavelet. The transforms fully cover the frequency axis,
but the overlap between different orders is substantial
\label{fig:b3four}}
\end{figure}

Two points more: first, about normalization. \index{wavelet!normalization}
Although our scaling function $\phi(x)$ is normalized in the right way
(its integral is 1), the coefficient in the two-scale 
equation (\ref{twoscale} is different from the conventional one, 
and, as a result, the filter coefficients $h_k$ sum to 1, not to $\sqrt2$. 
The integral over the wavelet profile
is zero (this ensures that $\psi(x)$ is really a wavelet), but its norm
\mbox{$\int \psi^2(x)\,dx\approx 0.2345$}; normally it is chosen to be~1. 
What matters, 
really, is that it is not zero and does not diverge; welcome to the
wavelet world of free normalization.

Second, about initial data. We can recursively apply the transformation
rules (\ref{dyadsm}, \ref{dyadw}) only if we assume that the data $a_0(m)$ belongs
to the class of smooth functions (those obtained by convolution with the scaling 
function). So, if the raw data comes in ticking at grid points (regular times),
we should smooth it once before starting our transform chain. If the raw data is
given at points $x$ that do not coincide with the grid, the right solution is to
distribute  it to the grid $m$ with the scaling weights $\phi(x-m)$.
This procedure was christened 'extirpolation' (inverse interpolation) by
\index{extirpolation}
Press et al. (\cite{ESpress}); a strange fact is that N-body people 
extirpolate all the time in their codes, but nobody wants to use the term.

So, we have specified all the recipes needed to perform the \emph{\`a trous}
transform. Before we do that, we have to answer the question -- why? 
Bernard Jones demonstrated us how well orthogonal wavelet transforms work.
An orthogonal wavelet transform changes a signal (picture) of $N$ pixels
into exactly $N$ wavelet amplitudes, while an \emph{\`a trous} transform 
expands it into $N\times J$ pictures; why bother?

The reason is called \index{translational invariance}
'translational invariance'. As many of the wavelet 
amplitudes of an orthogonal transform do not have an exact home, then,
when shifting the data, these amplitudes change in strange ways. Sure, we
can always use them to reconstruct the shifted picture, but it makes no
sense to compare the wavelet amplitudes of the original and shifted pictures.
All the \emph{\`a trous} transforms, in the contrary, keep their amplitudes,
these move together with the grid. This is 'translational invariance', and it
is important in texture studies, where we want to see different scales of
a picture at exactly the same grid point. And cosmic texture is the 
main subject of this lecture.

A point to note -- Fourier transforms are not translation invariant, too.

\subsection{Multi-Dimensional \`A Trous}

\index{wavelet transform!multi-dimensional}
All the above discussion was devoted to one-dimensional wavelets.
This is customary in wavelet literature, as the step into 
multi-dimensions is simple -- we form direct products of independent
one-dimensional wavelets, one for every coordinate. This has been the
main approach up to now, although it does not work well everywhere.
An important example is a sphere, where special spherical wavelets
have to be constructed (I suppose that these will be explained in the
CMB-lectures).

So, two-dimensional wavelets are introduced by defining the 2-D scaling
function
\beq
\label{phi2}
\phi(x,y)=\phi(x)\phi(y)\;,
\eeq
and three-dimensional wavelets -- by the 3-D scaling function
\beq
\label{phi3}
\phi(x,y,z)=\phi(x)\phi(y)\phi(z)\;.
\eeq
A little bit extra care has to be taken to define wavelets; we have to
step into Fourier space for a while for that.
Recalling (\ref{psifour}), we have to write for two dimensions
\[
\hat{\psi}(\omega_1,\omega_2)=\hat{\phi}(\omega_1/2)\hat{\phi}(\omega_2/2)-
   \hat{\phi}(\omega_1)\hat{\phi}(\omega_2)
\]
(the direct products (\ref{phi2}, \ref{phi3}) look exactly the same in the
Fourier space). For coordinate space, it gives
\[
\psi(x,y)=4\phi(2x)\phi(2y)-\phi(x)\phi(y)\;,
\]
and for three dimensions, respectively
\[
\psi(x,y,x)=8\phi(2x)\phi(2y)\phi(2z)-\phi(x)\phi(y)\phi(z)\;,
\]

\begin{figure}
\centering
\resizebox{0.6\textwidth}{!}{\includegraphics*{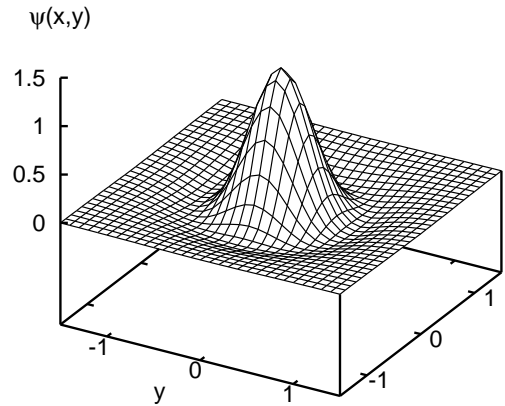}}
\caption{Two-dimensional $B_3$-associated wavelet \label{fig:b3-2d}}
\end{figure}

\begin{figure}
\centering
\resizebox{0.8\textwidth}{!}{\includegraphics*{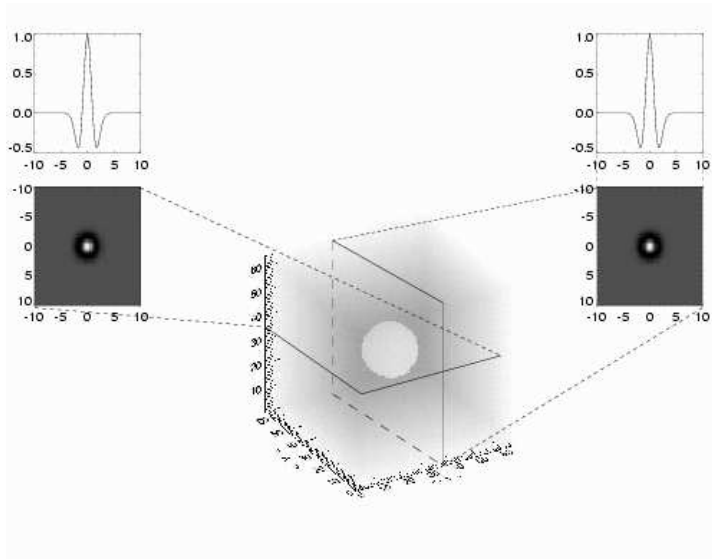}}
\caption{Three-dimensional $B_3$-associated wavelet \label{fig:b3-3d}}
\end{figure}

I show the $B_3$-associated wavelets in Figs.~\ref{fig:b3-2d} and 
\ref{fig:b3-3d}. Because of their definition, the wavelet profiles are
symmetric, but not isotropic, right? A big surprise is that both the
$B_3$ scaling functions and the wavelets are practically isotropic, as
the figures hint at. Let us define the isotropic part of the 2-D wavelet
as
\[
\overline{\psi}(r)=\frac1{2\pi}\int_0^{2\pi}\psi(r\cos\alpha,r\sin\alpha)\,d\alpha\;,
\]
and estimate the deviation from isotropy by
\[
\epsilon=\int_{-2}^2\int_{-2}^2|\psi(x,y)-\overline{\psi}(\sqrt{x^2+y^2})|\,dx\,dy\;.
\]
Comparing $\epsilon$ with the integral over the absolute value of our wavelet
itself (about 4/9), we find that the difference is about 2\%. For three 
dimensions, the difference is a bit larger, up to 5\%.

This isotropy is important for practical applications; it means that our
choice of specific coordinate directions does not influence the results
we get.

\begin{figure}
\centering
  \resizebox{!}{.9\textheight}{\includegraphics*{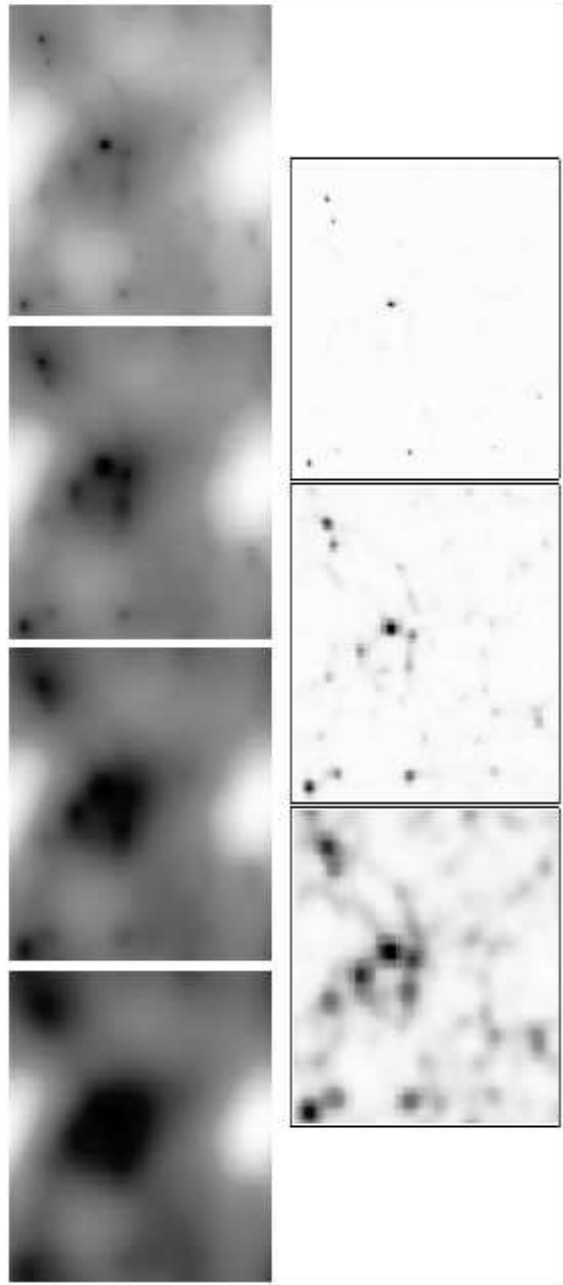}}
  \caption{\'A trous potential (slices) for a 
	N-body model (linear scale) \label{fig:pot}}
\end{figure}

And, as an example, I show a sequence of transforms in Fig.~\ref{fig:pot}.
Those are slices of a 3-D $B_3$-associated \emph{\`a trous} transform
sequence for the gravitational potential image of a N-body simulation.
The data slice is at the upper left; the left column shows the
scaling solutions, higher orders up. The right column shows
the wavelet amplitudes; the data can be restored by taking the lower left
image, and by adding to it all the wavelet images from the right column.
Simple, is it?

\section{Cosmological Densities}

In cosmology, the matter density distribution is a very important quantity that,
first, determines the future dynamics of structure, and, second, may carry
traces of the very early universe (initial conditions). We, however, cannot
observe it. The data we get, after enormous effort, gives us galaxy positions
in (redshift) space; even if we learn to associate a proper piece of dark
matter with a particular galaxy, we have a point process, not a continuous
density field.

There are several ways to deal with that. First, we have to introduce a 
coordinate grid; an extra discrete entity, but necessary when using
continuous fields in practical computations.

\subsection{N-Body Densities}

A similar density estimation problem arises in N-body calculations.
The fundamental building blocks there are mass points; these points are
moved around every timestep, and in order to get the accelerations
for the next timestep, the positions of the points have to be translated to
the mass density on the underlying grid.

\index{density!assigment!NGP}
The simplest density assignment scheme is called NGP (nearest grid point),
where the grid coordinate $i$ is found by rounding the point coordinate $x$
(to keep formulae simple, consider one-dimensional case and grid step one;
generalization is trivial). Thus, the NGP assignment law is
$i=\mathrm{floor}(x+0.5)$. This scheme is pretty rough, but I have seen people
using an even worse scheme, $i=\mathrm{floor}(x)$. 

\index{density!assigment!CIC}
A bit more complex scheme is CIC (cloud in cell) where the mass point is dressed
in a cubic cloud of the size of the grid cell, and vertices's are dressed in a 
similar region of influence. The part of the cloud that intersects this
region of influence is assigned to the vertex. Sounds complicated, but as the
mass weights are obtained by integration over the constant-density cloud, it
is, in fact, only linear extirpolation (in 3-D, three linear).

\index{density!assigment!TSC}
The most complex single-cloud scheme used is TSC (triangular shaped cloud),
where the density of the cloud changes linearly from a maximum in the centre,
to zero at the borders. The mass integration needed ensures that this scheme 
is quadratic extirpolation.

Note that the last two mass assignment schemes are, if fact, centred $B$-spline
assignments -- the vertex region is $B_0$, the CIC density is $B_0$, too,
and the TSC cloud is $B_1$. The weights are obtained by convolution of those
density profiles with the vertex profile, so, finally, NGP is a $B_0$
extirpolation scheme (no density law to be convolved with), CIC becomes
a $B_1$ extirpolation scheme and TSC -- a $B_2$ scheme.

Today's mass assignment schemes are mostly adaptive variations of those
listed above, where more dense grids are built in regions of higher density.
But the N-body mass assignment schemes share one property -- mass conservation.
No matter what scheme is used, the total mass assigned to all vertexes of the
grid is equal to the total mass of the mass points. No surprise in that --
box splines are interpolating (\ref{interpol}).

\subsection{Statistical Densities}

This class of density assignments (estimators) does not care about
mass conservation. The underlying assumption is that we observe a sample
of events that are governed by an underlying probability density,
and have to estimate this density. In cosmology, there really is no difference
between the two densities, spatial mass density and probability density.

The basic model of galaxy distribution adopted by cosmological statistics 
is that of the Cox point process. It says that, first, the universe is 
defined by a realization of a random process that ascribes a probability
density $\lambda(\mathbf{x})$ in space. Then, a Poisson point process gets
to work, populating the universe with galaxies, where the probability to
have a galaxy at $\mathbf{x}$ is given by the Poisson law with the
parameter $\lambda(\mathbf{x})$ fixed by the initial random-field realization.
Neat, right? And as we are dealing with random processes, no conservation 
is required.

Statisticians have worked seriously on probability density estimation
problem (see Silverman \cite{ESsilverman} for a review). 
The most popular density estimators are the kernel estimators:
\beq
\label{kernel}
\varrho_i=\sum_n K(x_n-i)
\eeq
(recall that $i$ are the grid coordinates and $x_n$ are the galaxy coordinates).
The kernel $K$ is a symmetric distribution:
\[
\int K(x)dx=1\;, \qquad \int xK(x)=0\;.
\]
Much work has been done on the choice of kernels, with the result that the
exact shape of the kernel does not matter much, but its width does.
\index{density!estimator!Epanechikov kernel}
The best kernel is said to be the Epanechikov kernel:
\beq
\label{epa}
K_E(\mathbf{x})=A(1-\mathbf{x}^2/R^2)\;,\quad \mathbf{x}^2\leq R^2,\; 
\quad 0\quad \mathrm{otherwise}\;.  
\eeq
(I wrote it for multidimensional case to stress that this is not a
direct-product, but an isotropic kernel; $A$ is, of course, a normalization 
constant).
\index{density!estimator!Gaussian kernel}
The Gaussian kernel comes close behind:
\beq
\label{Gauss}
K_G(x)=\frac1{\sqrt{2\pi}\sigma}\exp\left(-x^2/2\sigma^2\right)\;.
\eeq
This is the only kernel where the direct product is isotropic, too.
The ranking of kernels is done by deciding how close the estimated 
probability density $\tilde{f}(x)$ is to the true density $f(x)$,
by measuring the MSE (mean standard error):
\beq
\label{mse}
\mathrm{MSE}=E\left[\tilde{f}(x)-f(x)\right]^2
	=\mathrm{Var}\left(\tilde{f}(x)\right)
	+\left[\mathrm{Bias}\left(\tilde{f}(x)\right)\right]^2\;.
\eeq
Note that statisticians minimize the MSE, not only the variance, 
as cosmologists frequently tend to do. \index{MSE}
As usual in statistics, the results are asymptotic, true for a very big
number of galaxies $N$. As I have tested, in the usual cosmological
case where there are about 10 galaxies inside the Epanechikov kernel,
the total mass over all vertexes is only a half of the galaxy number $N$;
Epanechikov cheats. It starts working properly for about 100 points inside
the kernel. In this respect, the $B_3$ kernel we used above, is 
 is a very good candidate for a density estimation kernel.
It is smooth (meaning its Fourier transform
decays fast), and it is compact, no wide wings as the Gaussian
kernel has. And it is interpolating, guaranteeing that not a single galaxy
is lost.

Kernel density estimators allow a natural generalization for the case 
of extremely different density amplitudes and scales, as seen in cosmology.
Constant-width kernels tend to over-smooth the sharp peaks of the density,
if these exist. The solution is using adaptive kernels,
by varying the kernel width $h(\cdot)$ from place to place.
There are, basically, two different ways to do that.
\index{density!estimator!balloon} \index{density!estimator!scatter}
The balloon or scatter estimators say:
\beq
\label{balloon}
\varrho_i=\sum_n K\left(\frac{x_n-i}{h(i)}\right)\;;
\eeq
here the kernels sit on the grid points $i$.
\index{density!estimator!sample point} \index{density!estimator!scatter}
\index{density!estimator!sandbox} 
The second type of estimators is called sample point, sandbox, or gather
estimators:
\beq
\label{gather}
\varrho_i=\sum_n K\left(\frac{x_n-i}{h(x_n)}\right)\;.
\eeq
Here the kernel width depends on the sample point.
The most difficult problem for adaptive kernels
is how to choose the right kernel widths.
The usual way is to estimate the density with a constant kernel first, and
to select the adaptive kernel widths proportional to some fractional power
of the local density obtained in the first pass ($-1/5$ is a 
recommended choice).

Both estimators are used in cosmology (the terms scatter and gather come
from the SPH cosmological hydrodynamics codes). The lore says that the
balloon estimators (\ref{balloon}) work best in low-probability regions
(voids in cosmology), and the sandbox estimators -- where densities are
high.

\subsection{Equal-Mass Densities}

A popular density estimator is based on k-d trees. 
\index{density!estimator!kd-tree}
These trees are formed by recursive division of the sample space into two
equal-probability halves (having the same number of galaxies). It is
a spatial version of adaptive histograms (an equal number of events
per bin). Of course, k-d trees give more than just density estimates,
they also imprint a tree structure on (or reveal the structure of the
geometry of) the density field. An application of k-d trees for
estimating densities appeared in \url{astro-ph} during the school, and has
already been published by the time of writeup of the lecture
(Ascasibar \& Binney \cite{ESascasibar}).

\index{density!estimator!kNN kernel}
Another popular equal-mass density estimators are kNN ($k$ nearest
neighbours) kernels. The name speaks for itself -- the local kernel
size is chosen to include $k$ particles in the kernel volume. This estimator
uses isotropic kernels.

The SPH gather algorithm uses, in fact, the kNN ideology. There is a
separate free density estimation tool based on that algorithm
('\texttt{smooth}'), written by Joachim Stadel and available from the
Washington University N-body shop\footnote{\url
http://www-hpcc.astro.washington.edu/}. 
Try it; the only problems are that
you have to present the input data in the 'tipsy' format, and that you get
the densities at particle positions, not on a grid. Should be easy
to modify, if necessary.

\subsection{Wavelet Denoising}

\index{wavelet!denoising}
Wavelet denoising is a popular image processing methodology. 
The basic assumption is that noise in an image is present at all scales.
Once we accept that assumption, the way to proceed is clear:
decompose the image into separate scales (wavelet orders;
orthogonal wavelet transforms are the best here), estimate the noise at each 
wavelet order, eliminate it somehow, and reconstruct the image.

This course of actions includes two difficult points -- first,
estimating the noise. The properties of the basically unknown noise are,
ahem, unknown, and we have to make assumptions about them.
Gaussian and Poisson noise are the most popular assumptions;
this leaves us with the problem of relative noise amplitudes
(variances) between different wavelet orders. A popular method is
to model the noise. Modelling is started by assuming that all the
first-order wavelet data is noise (interesting, is it?) and processing
that for the noise variance. After that, noise of that variance is modelled,
wavelet transformed, and its properties found for every wavelet order.
After that, we face a a common decision theory problem, at which $p$-value
have to set the noise limit? If we cut the noise at too low amplitude, we
leave much of it in the final image, and if we take the cut too high, we
eliminate part of the real signal, too. Once we have selected that level,
we can quantify it in the terms of the limiting wavelet amplitude
$k\sigma_j$, where $\sigma^2_j$ is the modelled noise variance for
the level $j$.

The second problem is how to suppress the noisy amplitudes.
The first approach is called 'hard thresholding' \index{thresholding!hard}
and it is simple: the processed wavelet amplitudes $\tilde{w}_j$ are
\beq
\label{hard}
\tilde{w}_j=w_j\;,\quad \mbox{if}\quad {|w_j|>k\sigma_j}\;,\quad 0\quad \mbox{otherwise}\;.
\eeq
This thresholding leaves an empty trench around 0 in the wavelet amplitude
distribution.
\pagebreak

Another approach is 'soft thresholding': \index{thresholding!soft}
\beq
\label{soft}
\tilde{w}_j=w_j-\mbox{sgn}(w_j)\,k\sigma_j\;,\quad if\quad  {|w_j|>k\sigma_j}\;,\quad 0\quad \mbox{otherwise}\;.
\eeq
This thresholding takes out the same trench, first, but fills it up then,
diminishing all the remaining amplitudes. 

David Donoho, who was the first to introduce soft thresholding, has also
proposed an universal formula for the threshold level:
\[
k\sigma_j=\sqrt{2\log(n)}\,\sigma_j\;,
\]
where $n$ is the number of pixels in the image.
This level corresponds to $3\sigma$ for $n=90$, and to $4\sigma$ for $n=3000$.
Of course, astronomers complain -- 3000 pixels is a very small size for an
astronomical image, but $4\sigma$ is a very high cut-off level; we can
cut off much of the information in the image. Astronomical information is hard
to obtain, and we do not want to waste even a bit. So we better keep
our pictures noisy? Fortunately, more approaches to thresholding have
appeared in recent years; consult the new edition of the Starck and Murtagh
book (\cite{ESstarck06}).

\begin{figure}
\centering
\resizebox{0.45\textwidth}{!}{\includegraphics*{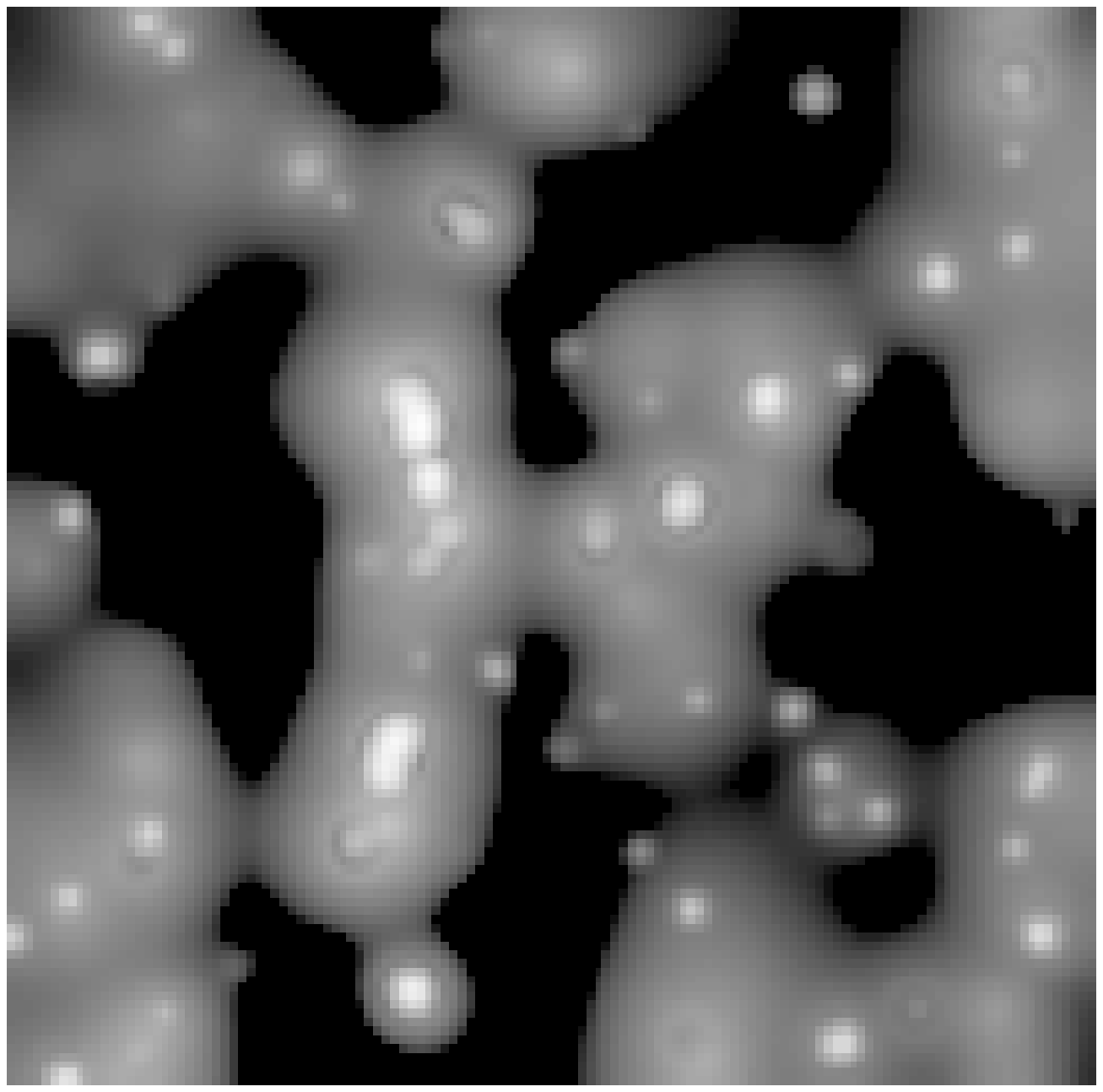}}
\resizebox{0.45\textwidth}{!}{\includegraphics*{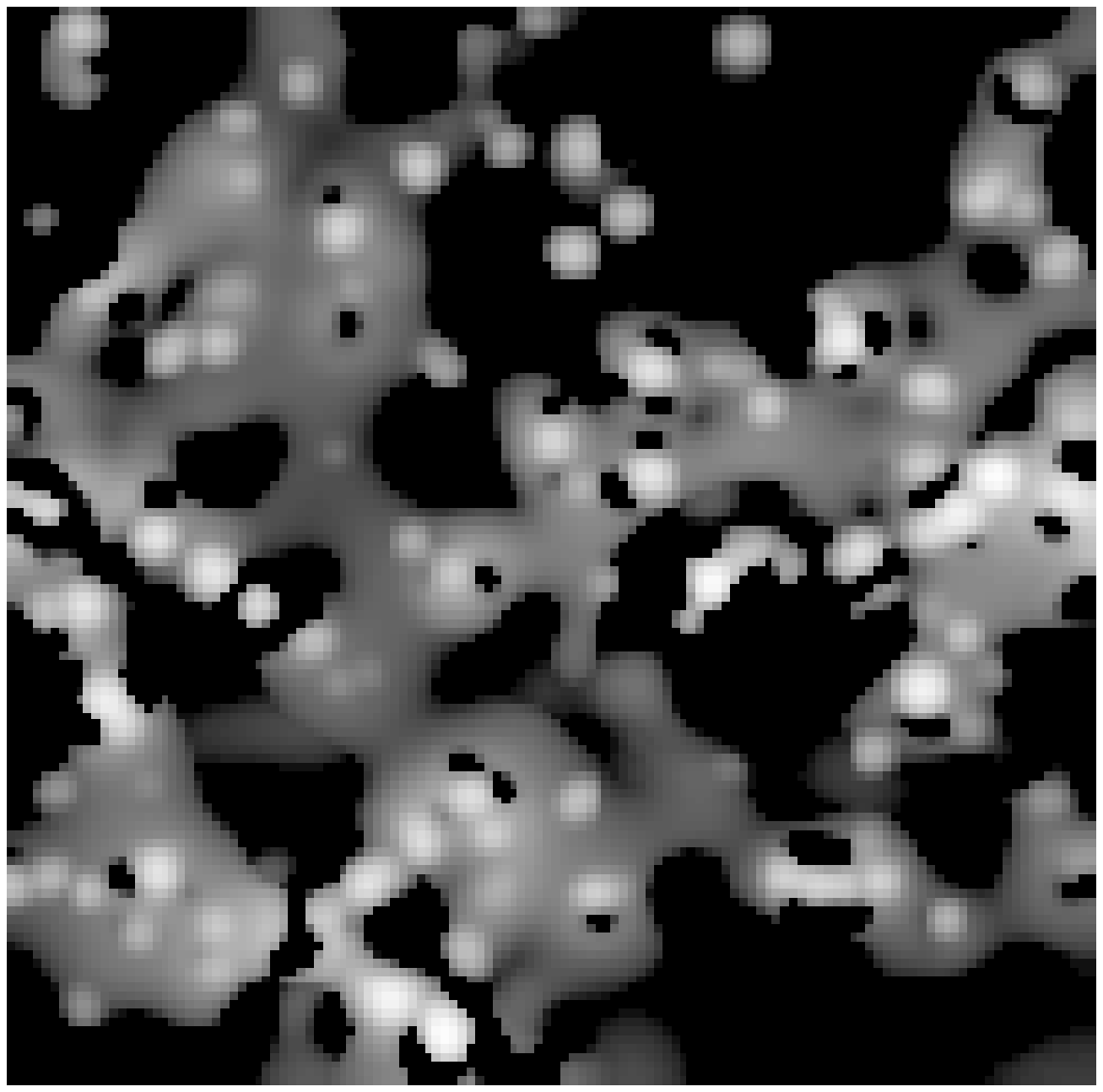}}\\
\caption{Wavelet denoising of a (model) galaxy distribution
(left -- a successful attempt, right -- over-denoising)\label{fig:3d_clean}}
\end{figure}

Anyway, wavelet denoising has met with resounding success in
image processing, no doubts about it. And image processing is an
industry these days, so the algorithms that are used are being
tested in practise every day. Now, image processing is 2-D business; wavelet
denoising of a 3-D (spatial) density is a completely different story.
The difference is that the density contrasts are much bigger in 3-D
than in 2-D -- there simple is more space for the signal to crowd in.
And as wavelets follow the details, they might easily over-amplify
the contrasts. I know, we have spent the last year trying to develop
a decent wavelet denoising algorithm for the galaxy distributions
(well, 'we' means mainly Jean-Luc Starck). Fig.~\ref{fig:3d_clean} shows
how the denoising might go awry (right panel), but shows at the same time
that good recipes are also possible (left panel). The denoising procedure
for the right panel has over-amplified the contrasts, and has generated
deep black (zero density) holes close to white high density
peaks. So, in order to do a decent denoising job, one has to be careful.

The details of the algorithm we used are too tedious to describe in full,
they can be found in Starck and Murtagh (\cite{ESstarck06})..
The main points are:
\begin{enumerate}
\item We used the \emph{\`a trous} algorithm, not an orthogonal one.
 The reason for that is that we needed to discriminate between the
positions of significant wavelet amplitudes (the multiresolution support) and
non-significant amplitudes at the last stage, and when speaking of positions, 
orthogonal wavelet transforms cannot be used.
\item We hard thresholded the solution and iterated it, reconstructing and
transforming again, to obtain a situation where the final significant 
wavelet coefficients would cover exactly the original multiresolution support.
\item Finally, we smooth the solution by imposing a smoothness constraint,
requiring that the sum of the absolute values of wavelet amplitudes of
a given order would be minimal, while keeping the corrected
amplitudes themselves within a given tolerance ($\sigma_j/2$) of the
original noisy ones. We consider here only the amplitudes in the
multiresolution support; this point required using a translation
invariant wavelet transform.
\end{enumerate}

\begin{figure}
\centering
\resizebox{0.35\textwidth}{!}{\includegraphics*{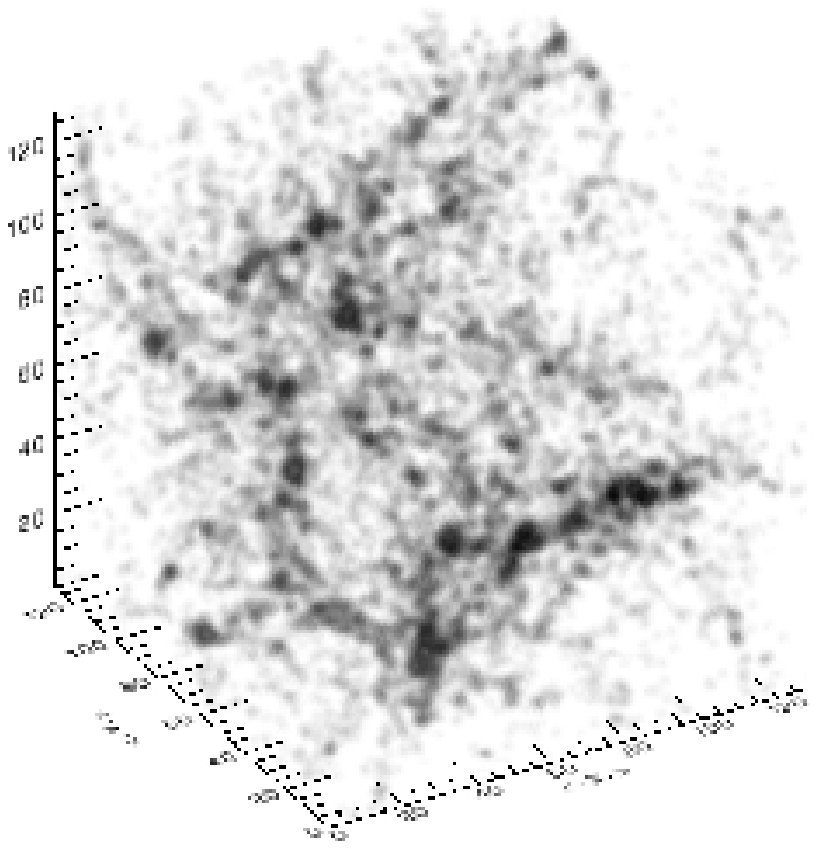}}
\resizebox{0.35\textwidth}{!}{\includegraphics*{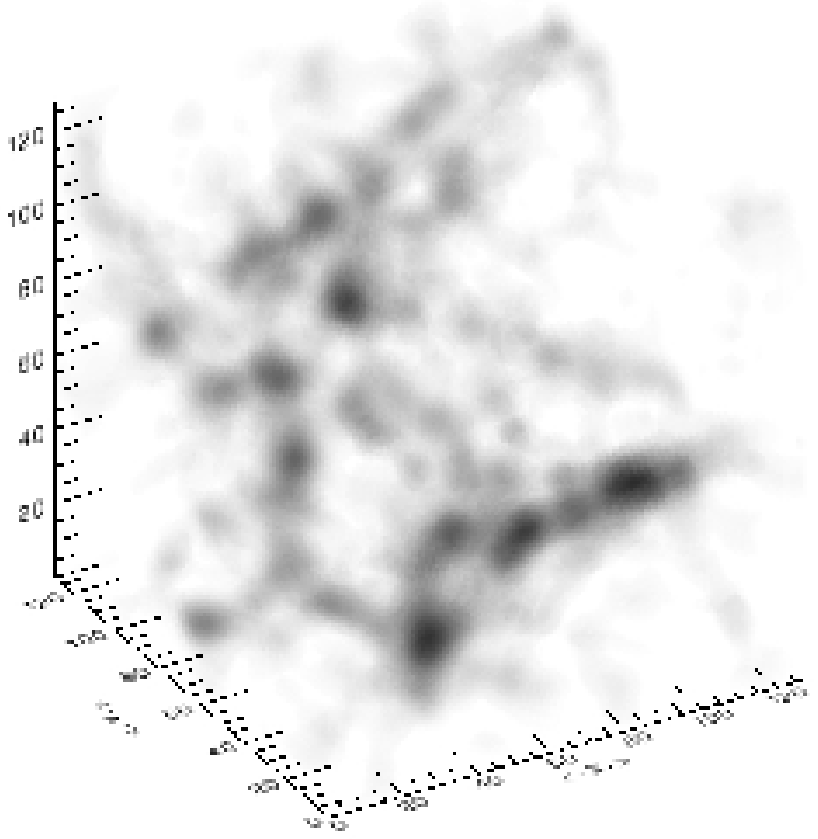}}\\
\resizebox{0.35\textwidth}{!}{\includegraphics*{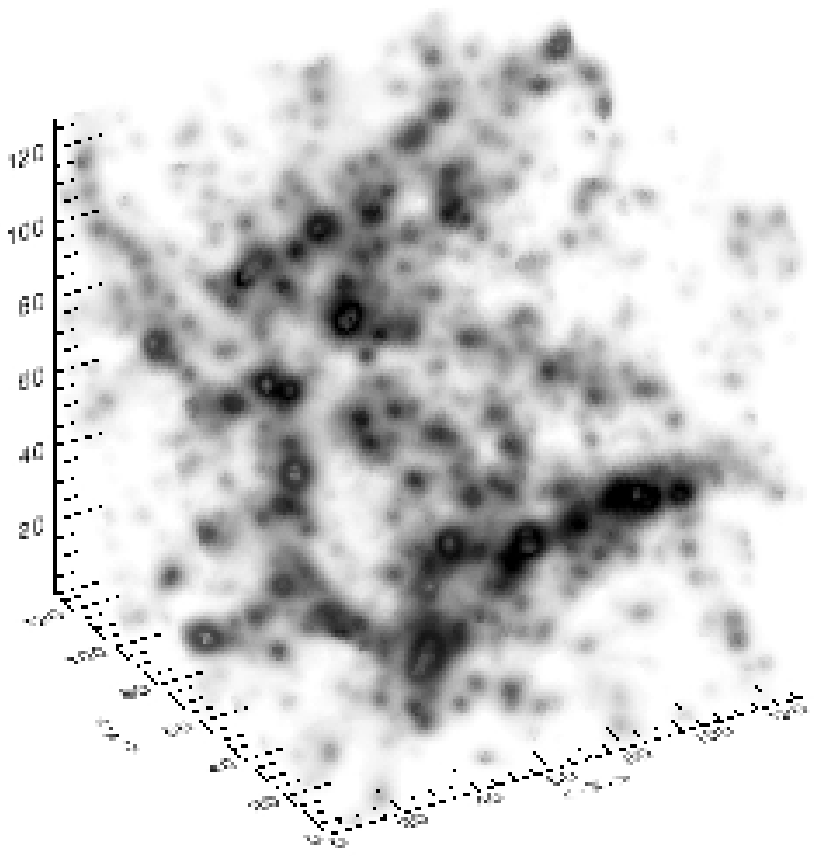}}\\
\caption{Density fields for a model galaxy distribution.
Left -- Gaussian $\sigma=1$\,Mpc/$h$ smoothing, middle -- wavelet
denoising, right -- Gaussian $\sigma=3$\,Mpc/$h$ smoothing\label{fig:3d_smooth}}
\end{figure}

Fig.~\ref{fig:3d_smooth} compares the results of our 3-D wavelet
denoising. We started with a 3-D galaxy distribution, applied our
algorithm to it, and, for comparison, built two Gaussian-smoothed
density distributions. It is clearly seen that the typical details, in
the case of Gaussian smoothing, are of uniform size, while the
wavelet-denoised density distribution is adaptive, showing details of
different scales.

\subsection{Multiscale Densities}

\index{multiscale!density}
We have made an implicit assumption during all this section,
namely, that a true density field exists. Is that so certain?
Our everyday experience tells us that it is. But look at the numbers
that stand behind this experience: even one cubic centimetre of air
has $6\times10^{23}/22.4\times 10^3\approx 3\times10^{19}$ particles.
In our surveys, one gigantic cosmological 'cubic centimetre' of 10\,Mpc size
contains about 10 galaxies. Can we speak about their true spatial density?

One answer is that we can, but there are regions where the density estimates
are extremely uncertain; statistics can tell us what the expected variances are.
Another answer is that even if there is a true density, it is not always a 
useful physical quantity, especially for the largest scales we study.
One of the reasons we measure the cosmological density field is to find its
state of evolution and traces of initial conditions in it. The theory of the
dynamics of perturbations in an expanding universe predicts that structure
evolves at different rates, slowly at large scales and much more rapidly
at galactic scales. Observations show that cosmological fields are multiscale
objects; the recently determined power spectra span scales from about
600\,Mpc/$h$ ($k=0.01\,h/\mathrm{Mpc}$ to 10\,Mpc/$h$ ($k=0.6\,h/\mathrm{Mpc}$.
Thus, should not these fields be studied in multiscale fashion, scale by scale?
A true adaptive density mixes effects from different scales and
scale separation could give us a cleaner look at the dynamics of large-scale
structure.

In case of our everyday densities, this separation of scales can
be done safely later, analyzing the true density. For galaxy distributions, 
it is wiser
to prescribe a scale (range) and to obtain that density directly from
the observed galaxy positions. One advantage is in accuracy, another
in that there are places (voids), where, e.g.,  small-scale densities
simply do not exist. 

\begin{figure}
\centering
\resizebox{0.7\textwidth}{!}{\includegraphics*{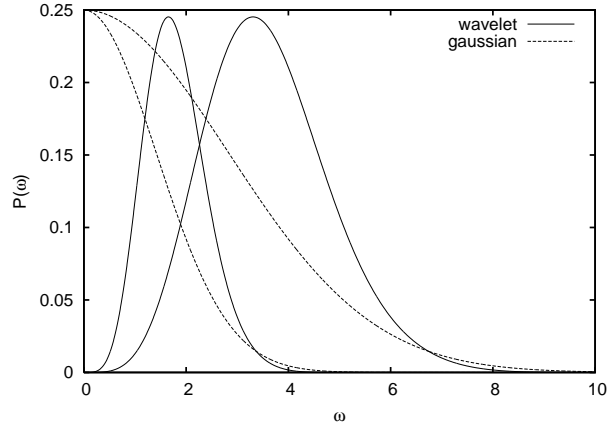}}
\caption{Power spectra for two neighbouring scales, the $B_3$-associated
\emph{\`a trous} bands (solid lines), and for Gaussian smoothing of
the same scales /(dotted lines). Only the positive half on the frequency
axis is shown  \label{fig:power}}
\end{figure}

And this is the point where the first part of the lecture (wavelets) connects
with the second part (density fields). The representation of the observed
density fields by a sum of the densities of different characteristic scales
(\ref{atrecon}) is just what we are looking for. True, there is a pretty
large frequency overlap between the neighbouring bands (Fig.~\ref{fig:power}
shows you their power spectra), but that is possibly the best we can do, while
keeping translational invariance. The figure shows also the power spectra of
Gaussians of the same scales that are sometimes used to select different
scales. As we see, Gaussian frequency bands are heavily correlated, the overlap
of the smaller frequency band with that of the higher one is total, and
that is natural -- smoothing destroys signals of high frequency, but it does 
not separate frequency bands. So Gaussians should not be used in this
business, but, alas, they frequently are.

\begin{figure}
\centering
  \resizebox{!}{.9\textheight}{\includegraphics*{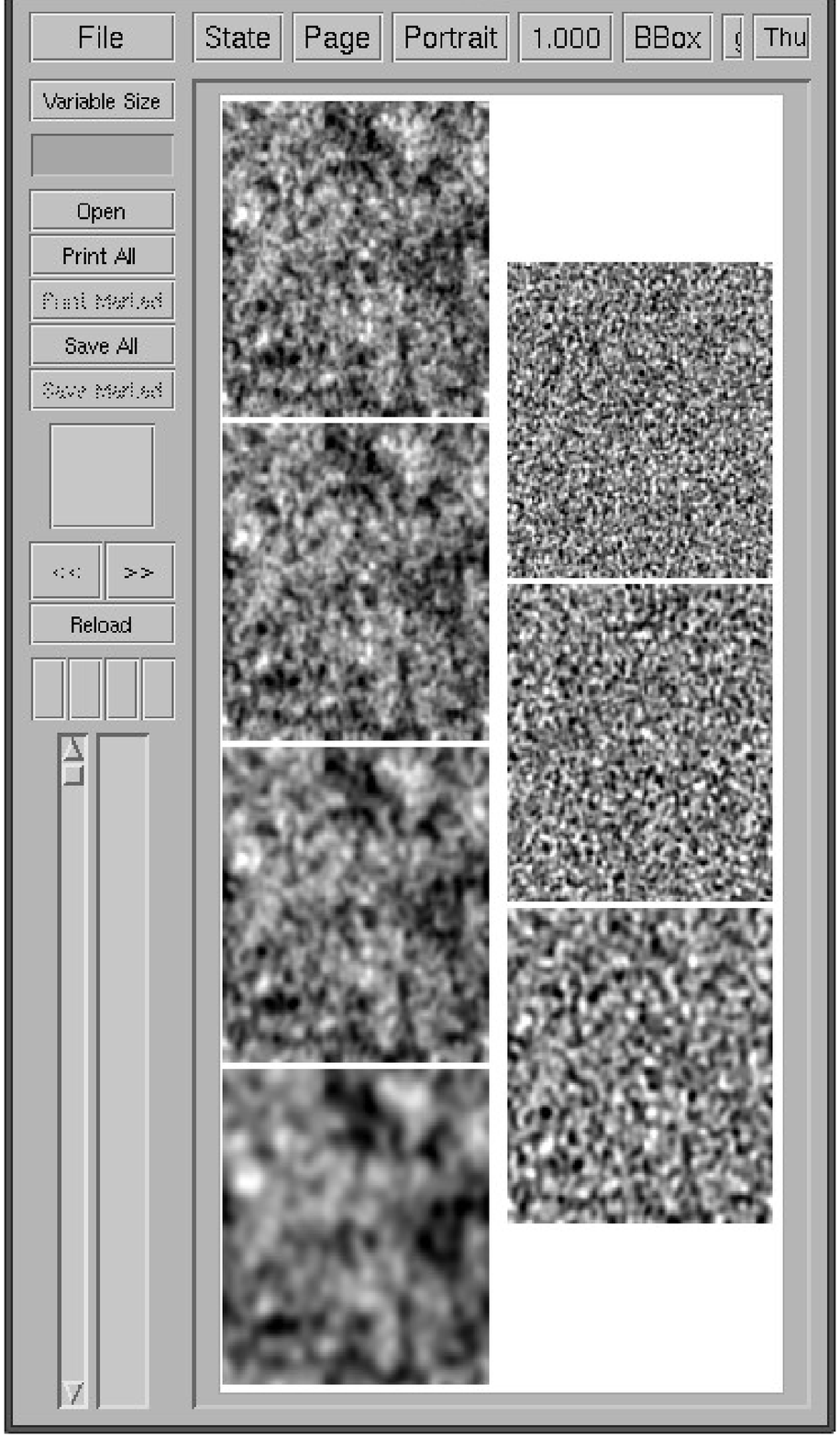}}
  \caption{\'A trous density (slices) of a Gaussian density field.
	(linear scale). The first orders (0--3 for the scaling solutions,
     1--3 for the wavelets)\label{fig:Ndens1}}
\end{figure}

\begin{figure}
\centering
  \resizebox{!}{.9\textheight}{\includegraphics*{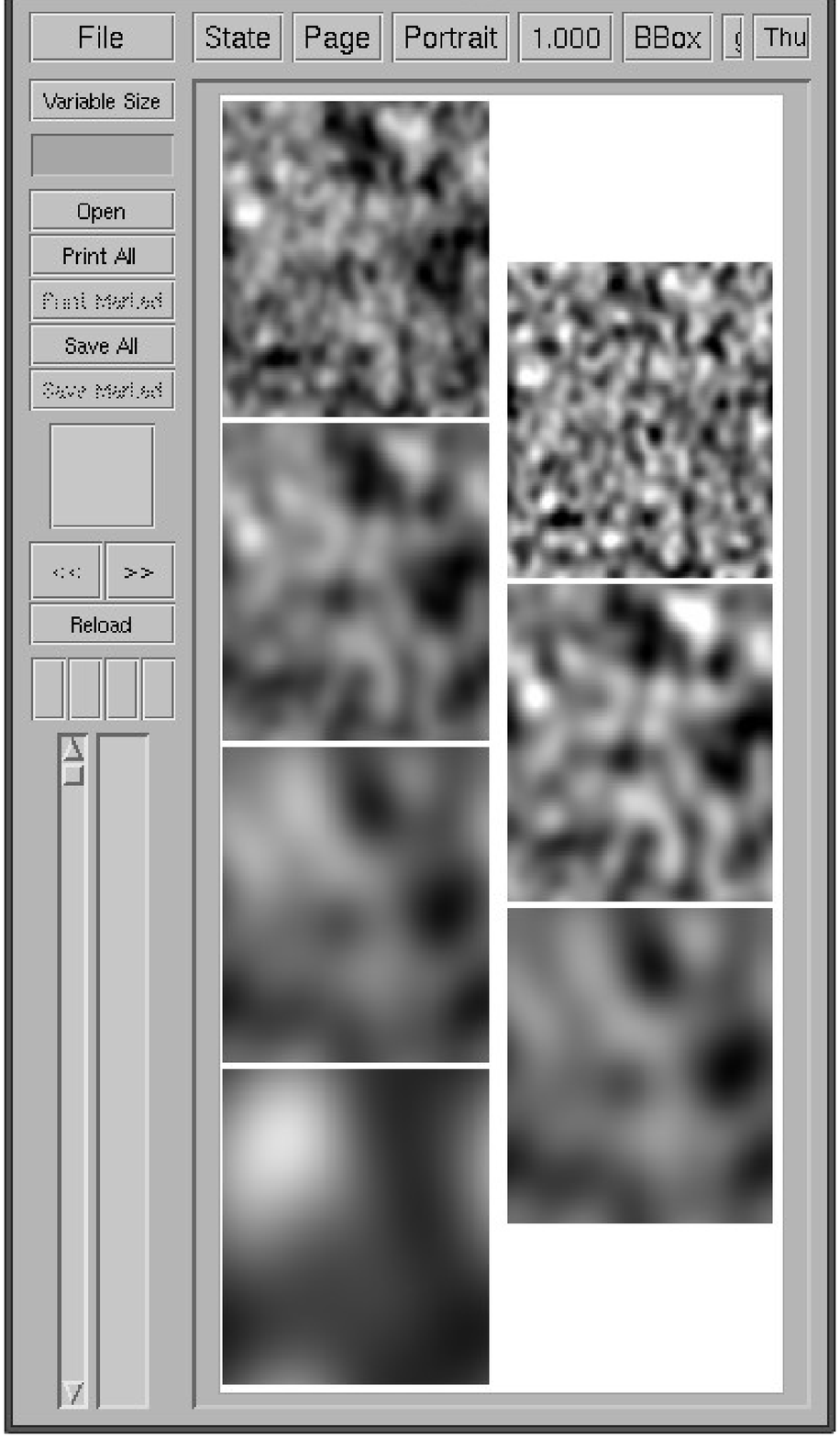}}
  \caption{\'A trous density (slices) of a Gaussian density field.
	(linear scale). The large-scale  orders (3--6 for the scaling solutions,
     4--6 for the wavelets)\label{fig:Ndens2}}
\end{figure}

Figs.~\ref{fig:Ndens1} and \ref{fig:Ndens2} show the \emph{\`a trous}
density slices for a Gaussian cube.
This is a $256^3$ realization of a Gaussian random field with a
power spectrum approximating that of our universe. In real universe,
the size of the cube would be 256\,Mpc/$h$. The slices are taken from
the same height; The images are in gray coding, black shows the densest regions.
As in Fig.~\ref{fig:pot}, the scaling solutions form the left
column and the wavelets -- the right column; transform orders grow downwards.
In height, the wavelet orders are placed between the scaling order that
produced them.
The scaling solution of order three is repeated in Fig.~\ref{fig:Ndens2}
to keep the scaling--wavelet alignment.
Enjoy.

\section{Minkowski Functionals}

Peter Coles explains in his lecture why it is useful to study morphology
of cosmological fields. In short, it is useful because it is sort
of a perpendicular approach to the usual moment methods. Our present
cosmological paradigm says that the initial perturbation field was a realization
of a Gaussian random field. The most direct test of that would be to
measure all $n$-point joint amplitude distributions, starting from the
1-point distribution. Well, we know that even this is not Gaussian, but we
know why (gravitational dynamics of a positive density field inevitably
skews this distribution), and we can model it. As cosmological densities
are pretty uncertain, the more uncertain are their many-point joint distributions.
So this direct check does not work, at least presently.

Another possibility to check for Gaussianity is to estimate higher order
correlation functions and spectra. For Gaussian realizations, odd-order
correlations and power spectra should be zero, and even-order moments
should be directly expressible via the second-order moments, the usual
two-point correlation function and the power spectrum. Their dynamical distortions
can also be modelled, and this is an active area of research.

Morphological studies provide an independent check of Gaussianity. Morphology
of (density) fields depends on all correlation functions at once, is
scale-dependent, but local, and can also be predicted and its change
caused by dynamical evolution can be modelled. This lecture is, finally,
about measuring morphology of cosmological fields.

\index{Minkowski functionals}\index{morphology}
An elegant description of morphological characteristics of
density fields is given by Minkowski functionals 
\cite{ESmecke94}. These functionals provide a complete family of
morphological measures -- all additive, motion invariant and
conditionally continuous\footnote{Note added during final revision: We submitted
recently a paper on multiscale Minkowski functionals, and the referee wondered
what does 'conditionally continuous' mean. So, now I know -- they are
continuous if the hypersurfaces are compact and convex, and we can
approximate any decent hypersurface by unions of such.}
functionals defined for any hypersurface
are  linear combinations of its Minkowski functionals.

The Minkowski functionals (MF for short) describe the morphology of
iso-density surfaces, and depend thus on 
the specific density level\footnote{In fact, Minkowski
functionals depend on a surface, that
is why they are called functionals (functions of functions). When we
specify the family of iso-density surfaces, the functionals will
depend, suddenly, only on a number, the value of the density level, and
are downgraded to simple functions, at least in cosmological applications.}.
Of course, when the original data are galaxy positions,
the procedure chosen to calculate densities (smoothing) will also
determine the result. The usual procedure used in this business is
to calculate kernel densities with wide Gaussian kernels; the recipes
say that the width of the kernel (standard deviation) should be either
the mean distance between galaxies or their correlation length, whichever
is larger. Although this produces nice smooth densities, the recipe
is bad, it destroys the texture of the density distribution; I shall show it 
later.

We shall use wavelets to produce densities, and shall look first at the
texture of a true (wavelet-denoised) density, and then at the
scale-dependent multiscale texture of the galaxy density distribution.
We could also start directly from galaxies themselves, as Minkowski
functionals can be defined for a point process,
decorating the points with spheres of the same radius, and studying
the morphology of the resulting surface. This approach does not refer to a 
density and we do not use it here. Although it is beautiful, too,
the basic model that it describes is a (constant-density)
Poisson process; a theory for
that case exists, and analytical expressions for Minkowski functionals
are known. Alas, as the galaxy distribution is strongly correlated, this
reference model does not help us much. The continuous density case has a
reference model, too, and that is a Gaussian random field, so this
is more useful.

For a $d$-dimensional space, one can find $d+1$ different Minkowski
functionals. We shall concentrate on usual 3-D space; for that,
the Minkowski functionals are defined as follows. Consider an
excursion set $F_{\phi_0}$ of a field $\phi(\mathbf{x})$ in 3-D (the set
of all points where density is larger than a given limit,
$\phi(\mathbf{x}\ge\phi_0$)). Then, the first
Minkowski functional (the volume functional) is the volume of 
this region (the excursion set):
\beq
\label{mf0}
V_0(\phi_0)=\int_{F_{\phi_0}}d^3x\;.
\eeq
The second MF is proportional to the surface area
of the boundary $\delta F_\phi$ of the excursion set:
\beq
\label{mf1}
V_1(\phi_0)=\frac16\int_{\delta F_{\phi_0}}dS(\mathbf{x})\;,
\eeq
(but not the area itself, notice the constant).
The third MF is proportional to the \index{integrated mean curvature}
integrated mean curvature
of the boundary:
\beq
\label{mf2}
V_2(\phi_0)=\frac1{6\pi}\int_{\delta F_{\phi_0}}
    \left(\frac1{R_1(\mathbf{x})}+\frac1{R_2(\mathbf{x})}\right)dS(\mathbf{x})\;,
\eeq
where $R_1(\mathbf{x})$ and $R_2(\mathbf{x})$ 
are the principal curvatures of the boundary.
The fourth Minkowski functional is proportional to the integrated
Gaussian curvature (the Euler characteristic) \index{Euler!characteristic}
of the boundary:
\beq
\label{mf3}
V_3(\phi_0)=\frac1{4\pi}\int_{\delta F_{\phi_0}}
    \frac1{R_1(\mathbf{x})R_2(\mathbf{x})}dS(\mathbf{x})\;.
\eeq
The last MF is simply related to the genus \index{genus}
that was the first morphological measure used in cosmology; 
all these papers bear titles containing the word 'topology'.
Well, the topological Euler characteristic $\chi$ for a surface in 3D
can be written as
\beq
\label{chisurf}
\chi=\frac1{2\pi}\int_S \kappa\,dS\;,
\eeq
where $\kappa$ is the Gaussian curvature, so
\beq
\label{chiv3}
V_3=\frac12\chi\;.
\eeq
Bear in mind, though, that the Euler characteristic (\ref{chisurf})
describes the topology of a given iso-density surface, not of
the full 3-D density distribution; the topology of the latter is, hopefully,
trivial.
\pagebreak

The first topology papers concentrated on the genus $G$ that is
similar to $V_3$:
\beq
\label{chigenus}
\chi=2(1-G)\;, \qquad V_3=1-G\;.
\eeq
The functional $V_3$ is a bit more comfortable to use -- it is additive,
while $G$ is not, and in the case our surface breaks up into several
isolated balls, $V_3$ is equal to the number of balls. If the excursion set
contains only a few isolated empty regions (bubbles), $V_3$ gives their
number. In a general case
\[
V_3=\#\mbox{-of-balls}+\#\mbox{-of-bubbles}-\#\mbox{-of-tunnels}\;,
\]
where only these tunnels that are open at both ends, count.

I have to warn you about a possible confusion with the genus relations
(\ref{chisurf}--\ref{chigenus}) -- the coefficient 2 (or 1/2) occupies
frequently a wrong position. The confusion is due to a fact that 
two topological characteristics can be defined for an excursion set -- 
one for its
surface, another for the set itself. The relation between these depends on
the dimensionality of the space; for 3-D the topological characteristic for
the excursion set is half of that for the surface, and if we mix them
up, our formulae become wrong. I know, we have published
a wrong formula, too (even twice), but the formulae are right in
our book \cite{ESmartsaar}. So, bear in mind that the
Minkowski functionals are calculated for surfaces, and use only the
relations above (\ref{chisurf}--\ref{chigenus}). When in doubt, consult
the classical paper by Mecke et al. \cite{ESmecke94}, and use the
Crofton's formula below (\ref{crofton}) for a single cubic cell -- it
gives you $V_3=1\Rightarrow G=0$¤.
 
\begin{figure}
\centering
\resizebox{0.4\textwidth}{!}{\includegraphics*{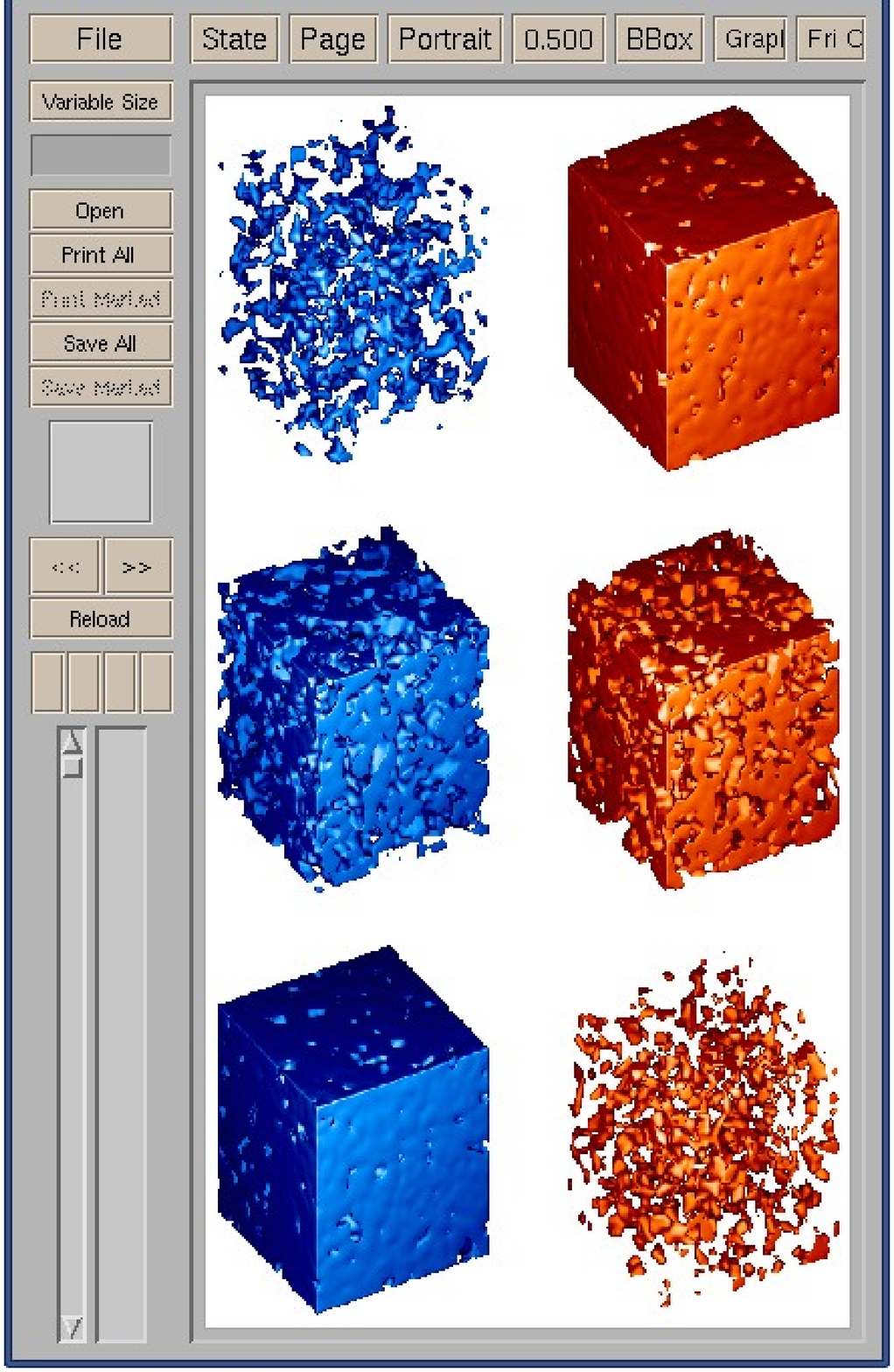}}
\hspace*{0.1\textwidth}
\resizebox{0.4\textwidth}{!}{\includegraphics*{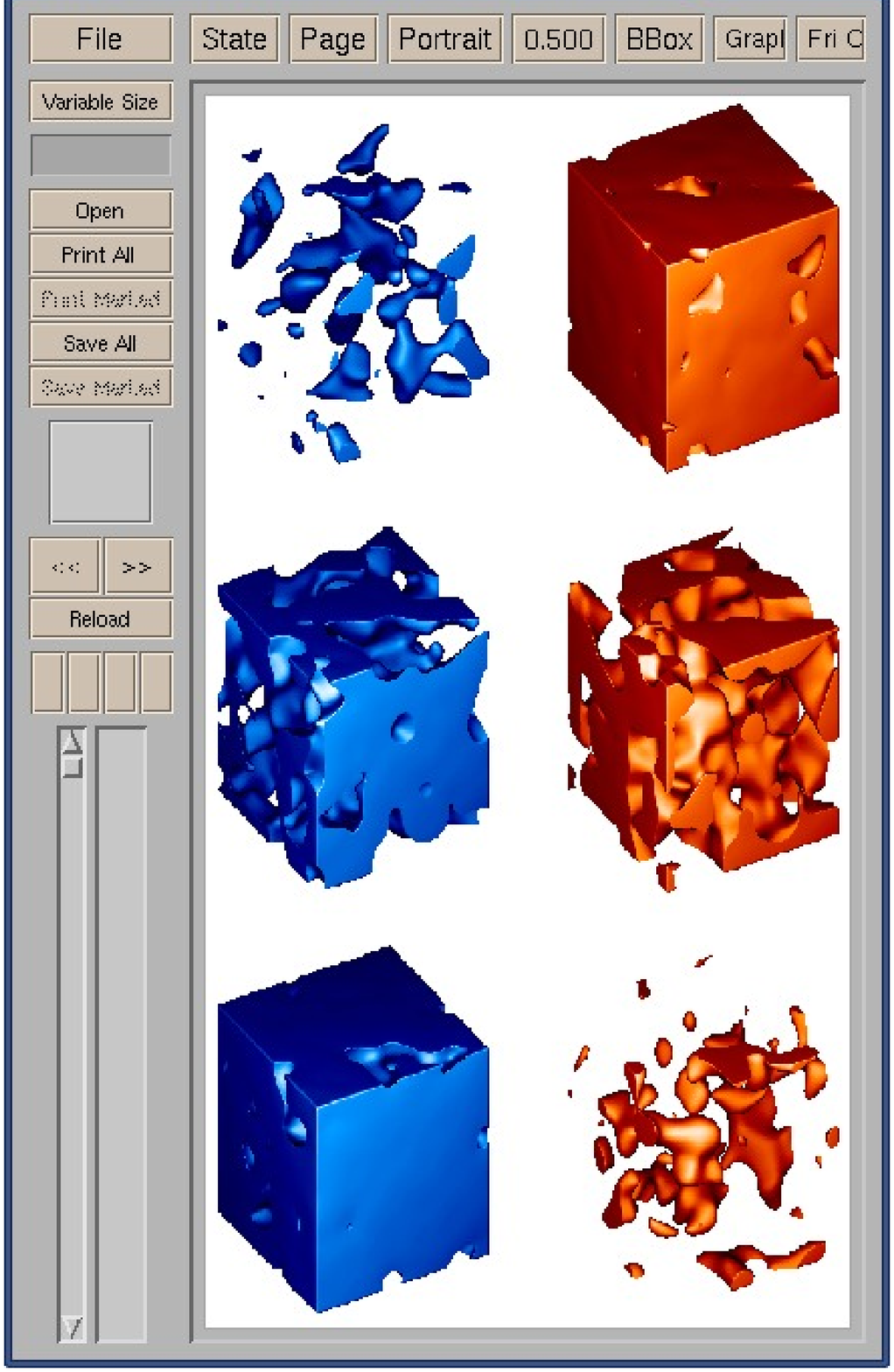}}\\
\caption{ A Gaussian cube of $256^3$ pixels for different Gaussian
smoothing filters. The left two columns show isodensity surfaces for 
$\sigma=3$ pixels, the right two columns -- for $\sigma=8$ pixels.
To better delineate isodensity surfaces, we show two sides of the
surface in column pairs, where the left column shows high-density regions, 
and the right column -- low-density regions for the same isodensity surface.
The rows are for constant volume fractions (7\%, 50\%, and 93\%), 
starting from below
\label{fig:gausscubes}}
\end{figure}

Fig.~\ref{fig:gausscubes}
shows a Gaussian cube (a realization of a Gaussian random process)
for two different smoothing widths (the left pair and the right pair
of columns, respectively), and for three volume fractions.
You can see that the solid figures inside the isodensity surface are
awash with handles, especially at the middle 50\% density level. Of course,
the larger the smoothing, the less the number of handles. You can also
see that Gaussian patterns \index{Gaussian!pattern}
are symmetric -- the filled regions are exact
lookalikes of the empty regions, for a symmetric change of volume fractions.

\begin{figure}
\centering
\resizebox{0.4\textwidth}{!}{\includegraphics*{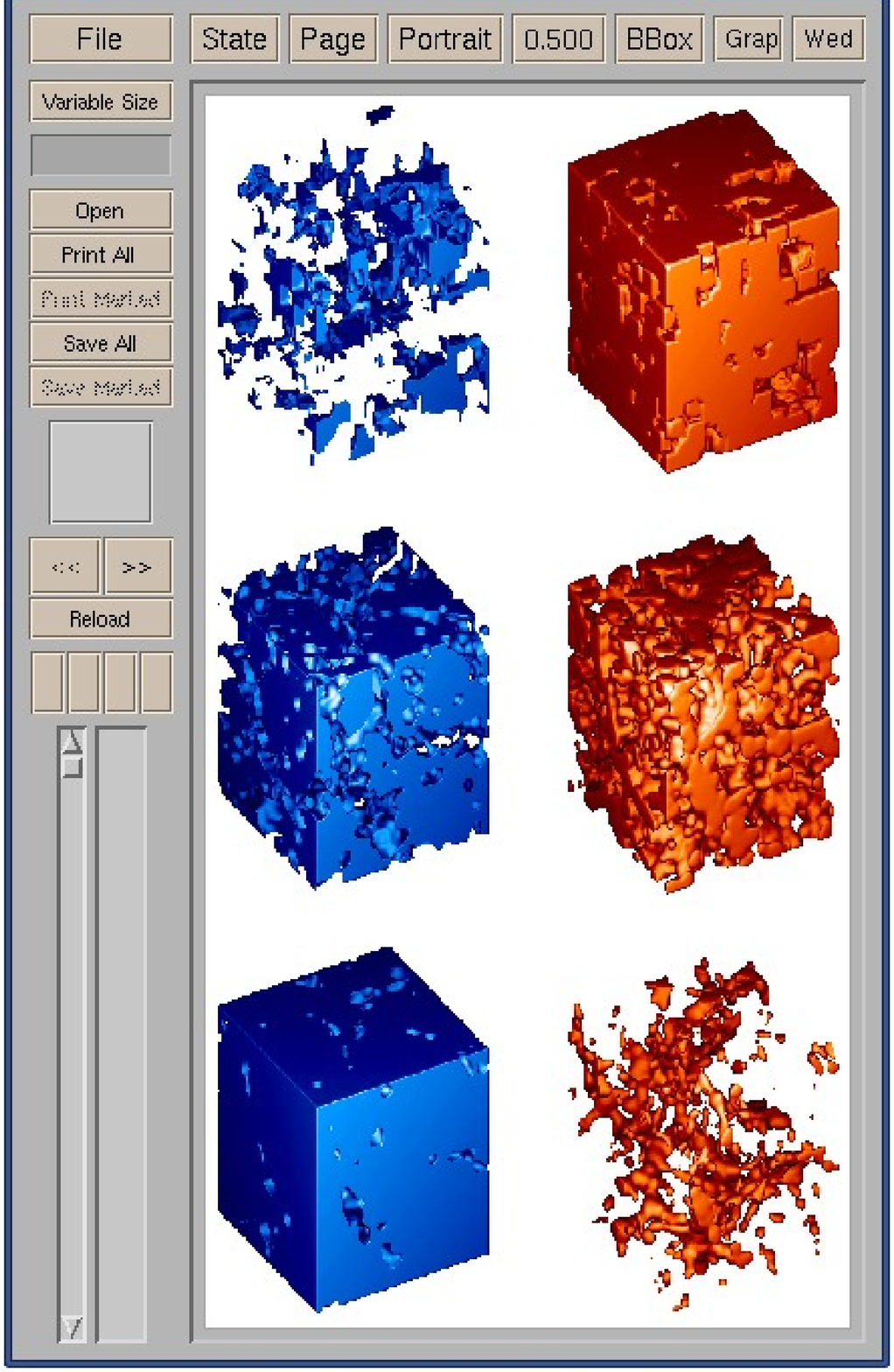}}\\
\caption{A galaxy sample ($60^3$ pixels) for a 3-pixel smoothing scale.
The left column shows high-density regions, 
and the right column -- low-density regions for the same isodensity surface.
The rows are for constant volume fractions (7\%, 50\%, and 93\%), 
starting from below
\label{fig:galcube}}
\end{figure}

Galaxy densities are more asymmetrical, as seen in Fig.~\ref{fig:galcube}.
This figure shows a model galaxy distribution from a N-body simulation, in
a smaller cube. The 50\% density volumes differ, showing asymmetry in the 
density
distribution, and the 5\% -- 95\% symmetry, evident for the Gaussian cube,
is not so perfect any more.

Instead of the functionals, their spatial densities $V_i$ are
frequently used:
\[
v_i(f)=V_i(f)/V\;, \quad i=0,\dots,3\;,
\]
where $V$ is the total sample volume. The densities allow us to
compare the morphology of different data samples.

\subsection{Labelling the Isodensity Surfaces}

The original argument of the functionals, the density level $\varrho_0$,
can have different amplitudes for different fields, and the functionals
are difficult to compare. Because of that, normalized arguments are
usually used; the simplest one is the volume fraction $f_v$, the ratio
\index{fraction!volume}
of the volume \emph{outside} of the excursion set to the 
total volume of the region
where the density is defined (the higher the density level, the closer
this ratio is to 1). Another, similar argument is the mass
fraction $f_m$, which is very useful for real, positive density fields,
\index{fraction!mass}
but is cumbersome to apply for realizations of Gaussian fields,
where the density may be negative. But when we describe the morphology
of single objects (superclusters, say), the mass fraction is the most
natural argument. It is also defined to approach 1 for the highest density
levels (and for the smallest masses inside the isodensity surface).

The most widely used argument in this business
is the Gaussianized volume fraction $\nu$, defined as
\index{fraction!volume!Gaussianized}
\beq
\label{nu}
f_v=\frac1{\sqrt{2\pi}}\int_\nu^\infty\exp(-t^2/2)\,dt\;.
\eeq
This argument was introduced
already in the first topology paper by Gott \cite{ESgott86}, 
in order to eliminate the first trivial
effect of gravitational clustering, the deviation of the 1-point pdf
from the (supposedly) Gaussian initial pdf. Notice that
using this argument, the first Minkowski functional is trivially
Gaussian by definition. 

\begin{figure}
\centering
  \resizebox{.45\textwidth}{!}{\includegraphics*{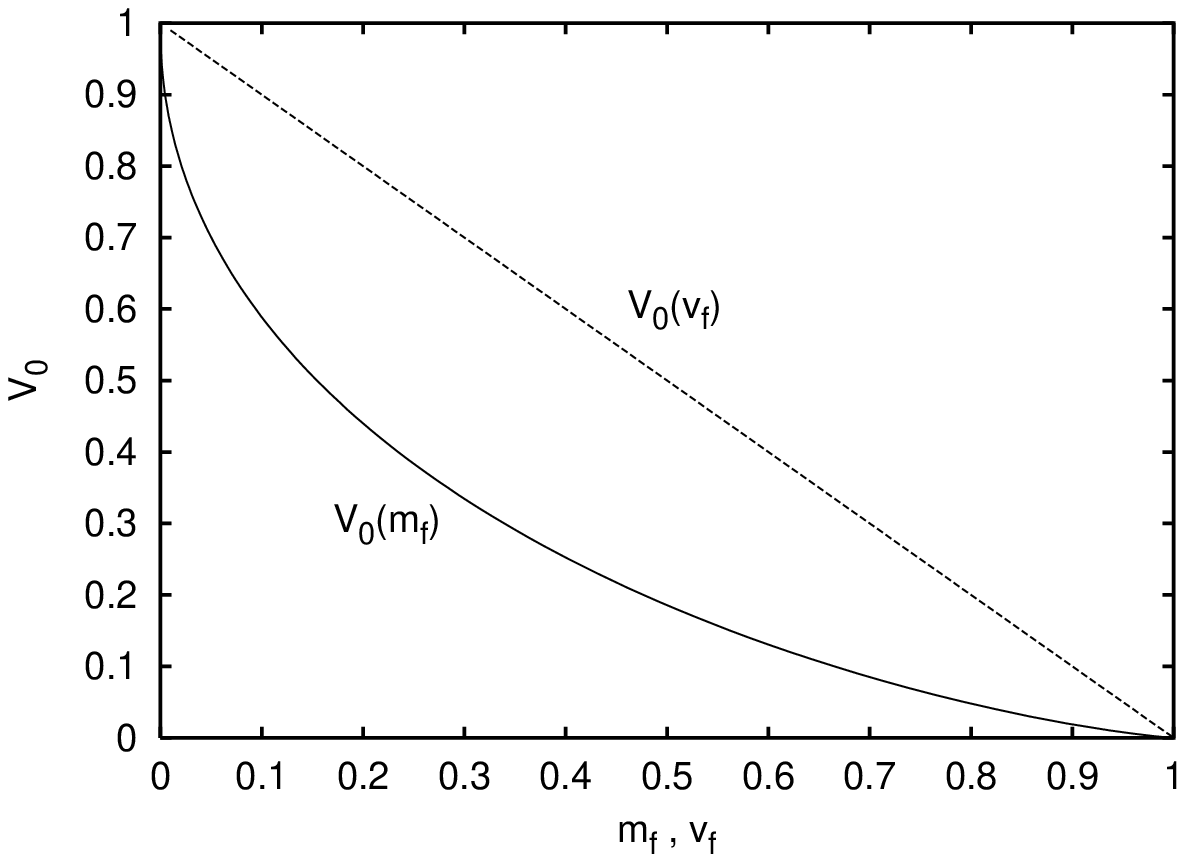}}
  \resizebox{.45\textwidth}{!}{\includegraphics*{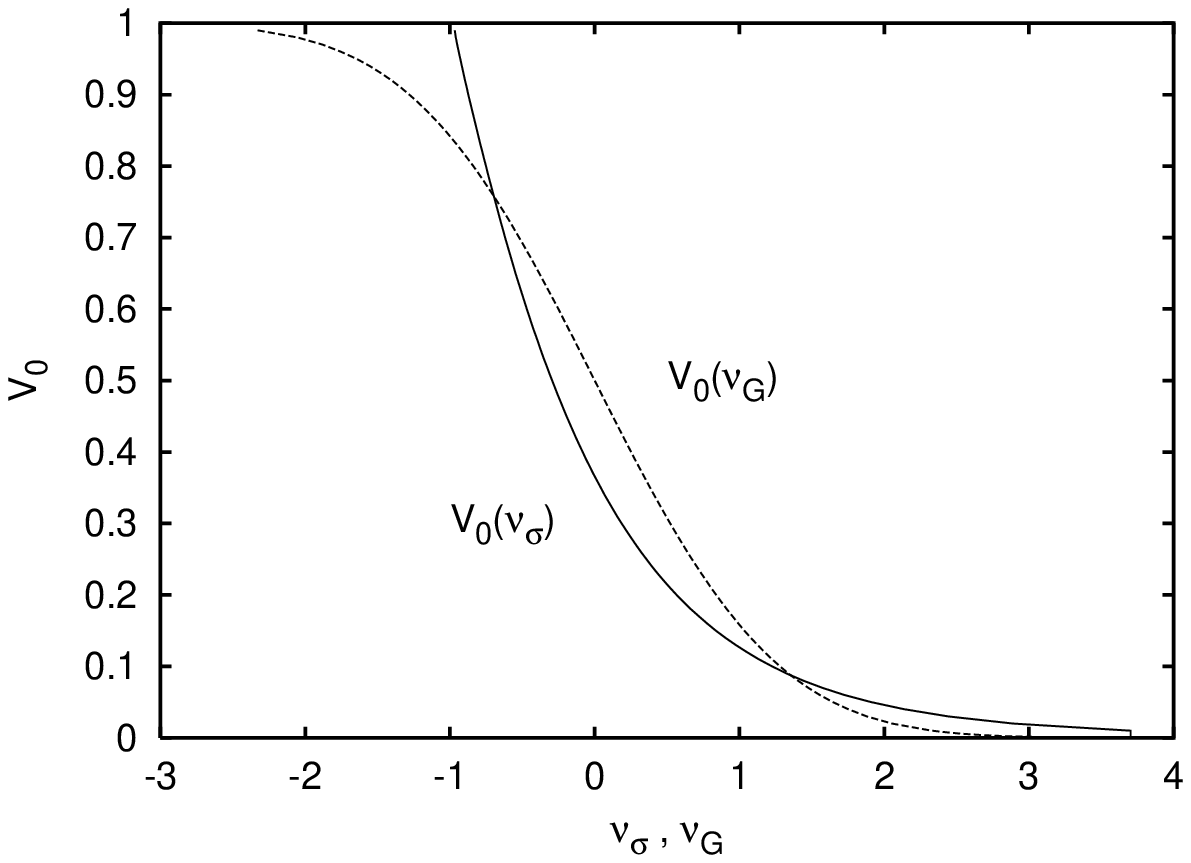}}\\
  \resizebox{.45\textwidth}{!}{\includegraphics*{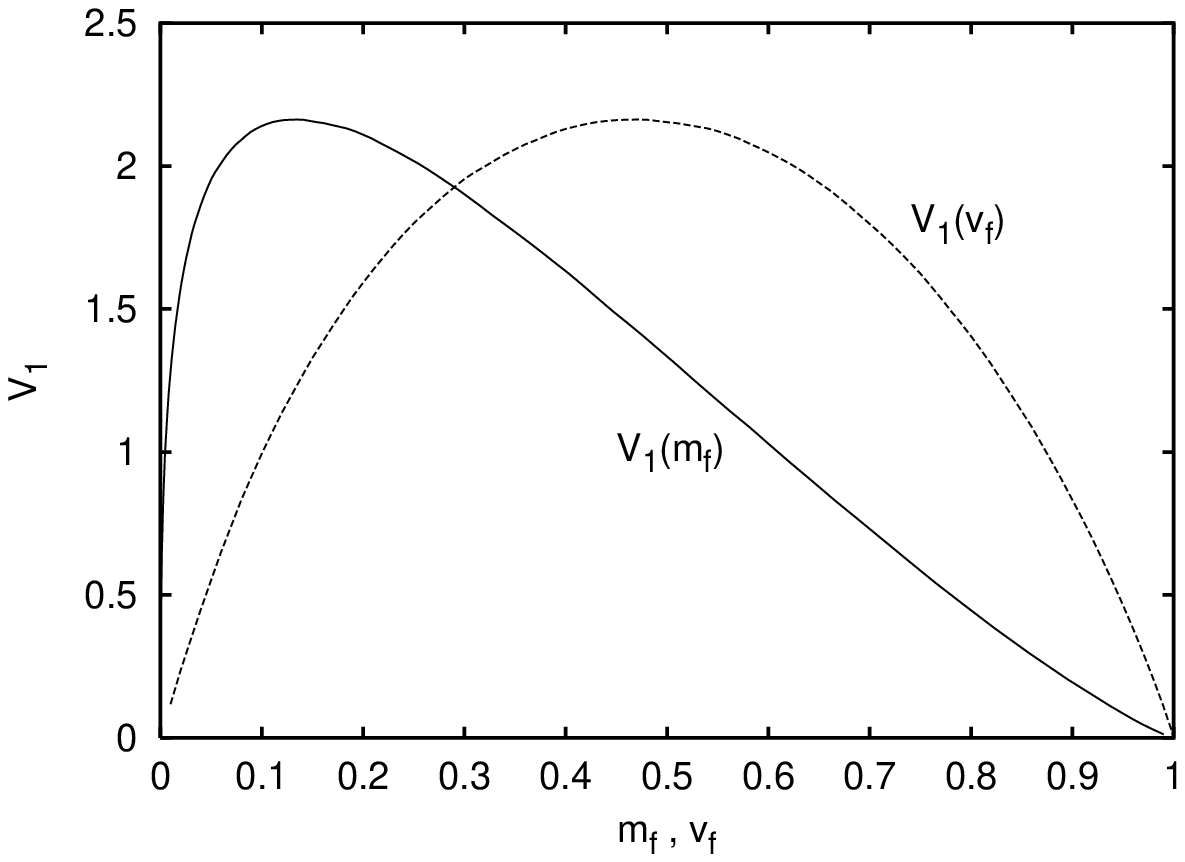}}
  \resizebox{.45\textwidth}{!}{\includegraphics*{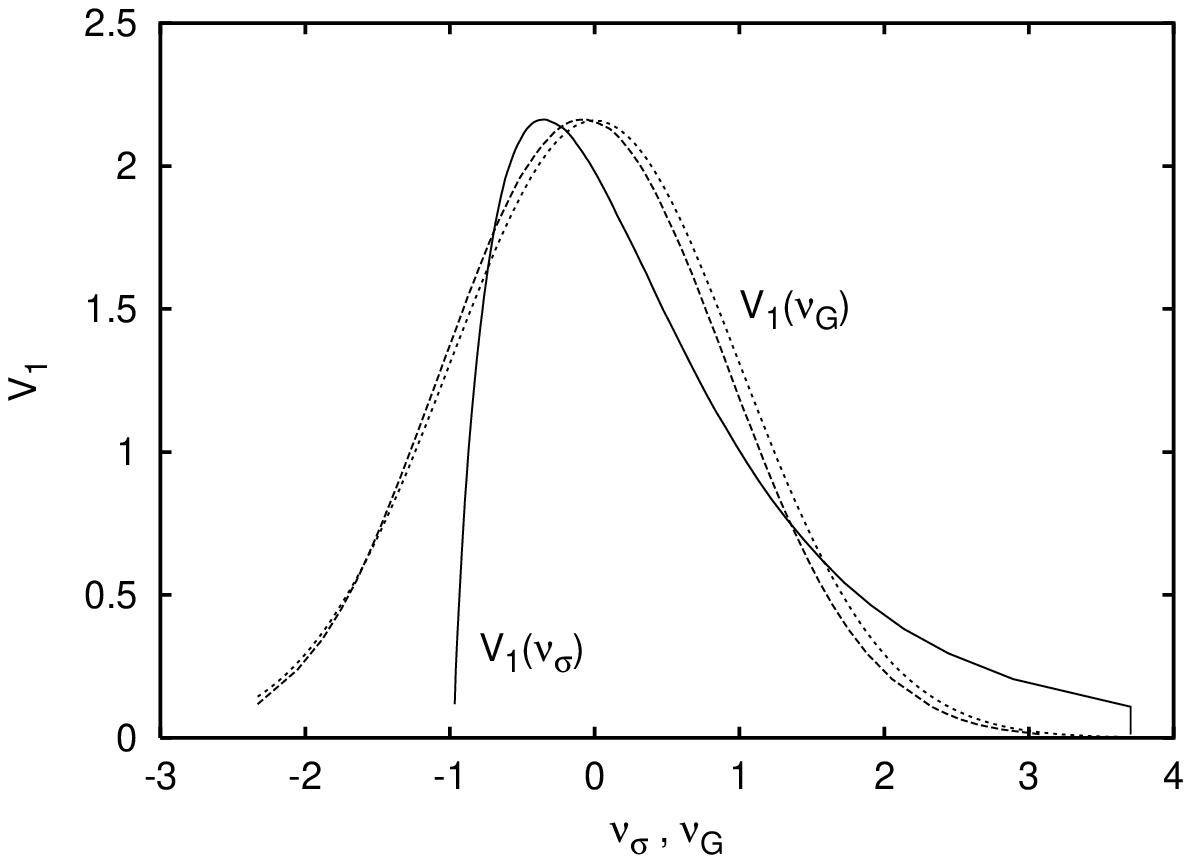}}\\
\caption{The first two Minkowski functionals for N-body model galaxies.
Here $v_f$ is the volume fraction, $m_f$ -- the mass fraction, 
$\nu_\sigma$ -- the normalized volume fraction, and $\nu_G\equiv\nu$ --
the Gaussianized volume fraction from (\ref{nu}).
The dotted line in the right panels
shows the predicted Minkowski functionals for a Gaussian random field MF
\label{fig:MF1_GIF}}
\end{figure}

For a Gaussian random field, $\nu$ is the density deviation from the
mean, divided by the standard deviation. We \index{fraction!volume!normalized}
can define a similar argument for any field:
\[
\nu_\sigma=\frac{\varrho-\overline{\varrho}}{\sigma(\varrho)}\;.
\]

\begin{figure}
\centering
  \resizebox{.45\textwidth}{!}{\includegraphics*{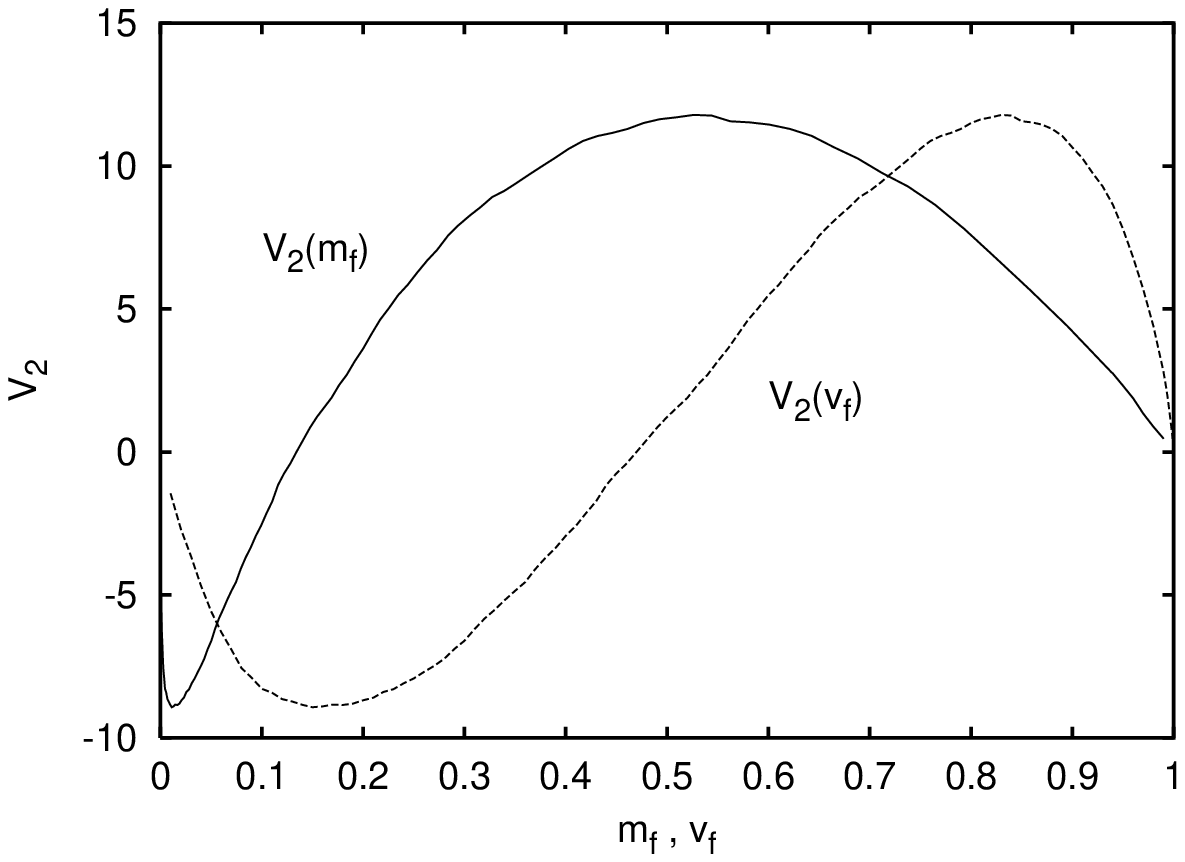}}
  \resizebox{.45\textwidth}{!}{\includegraphics*{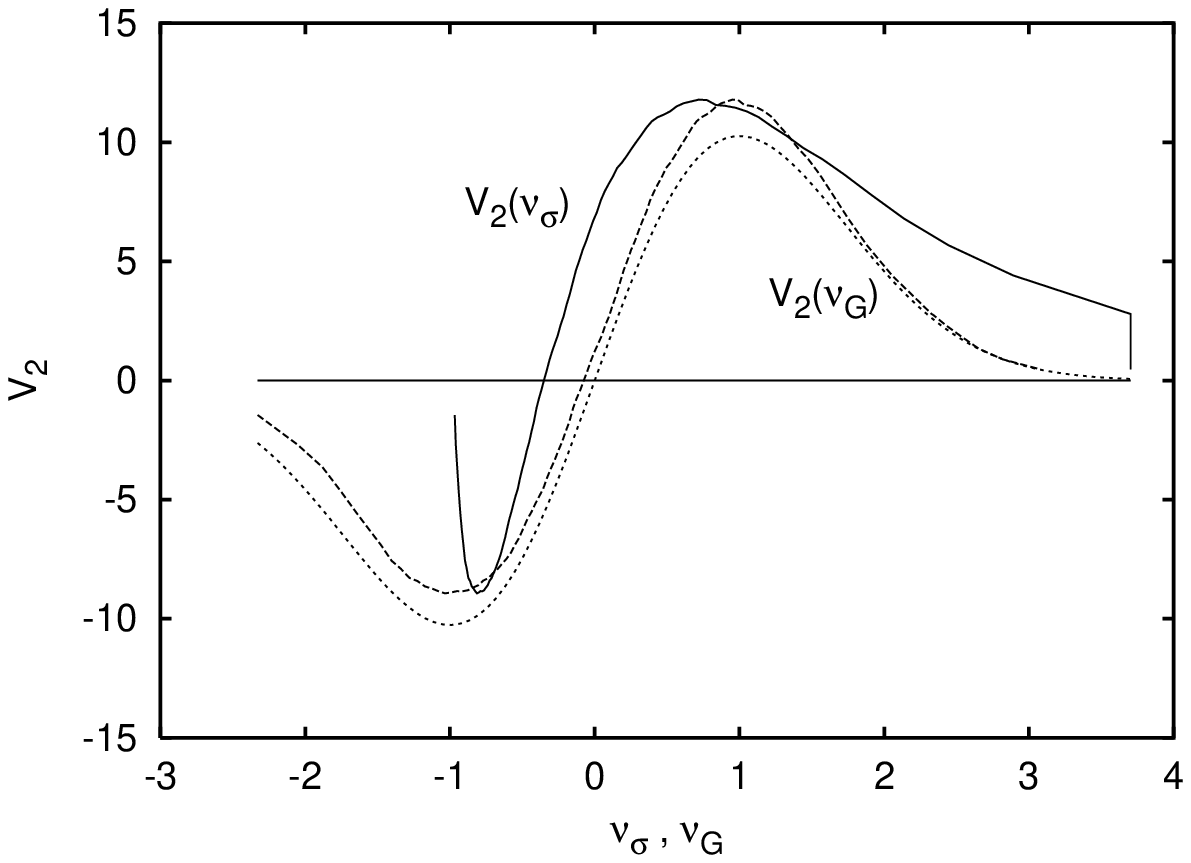}}\\
  \resizebox{.45\textwidth}{!}{\includegraphics*{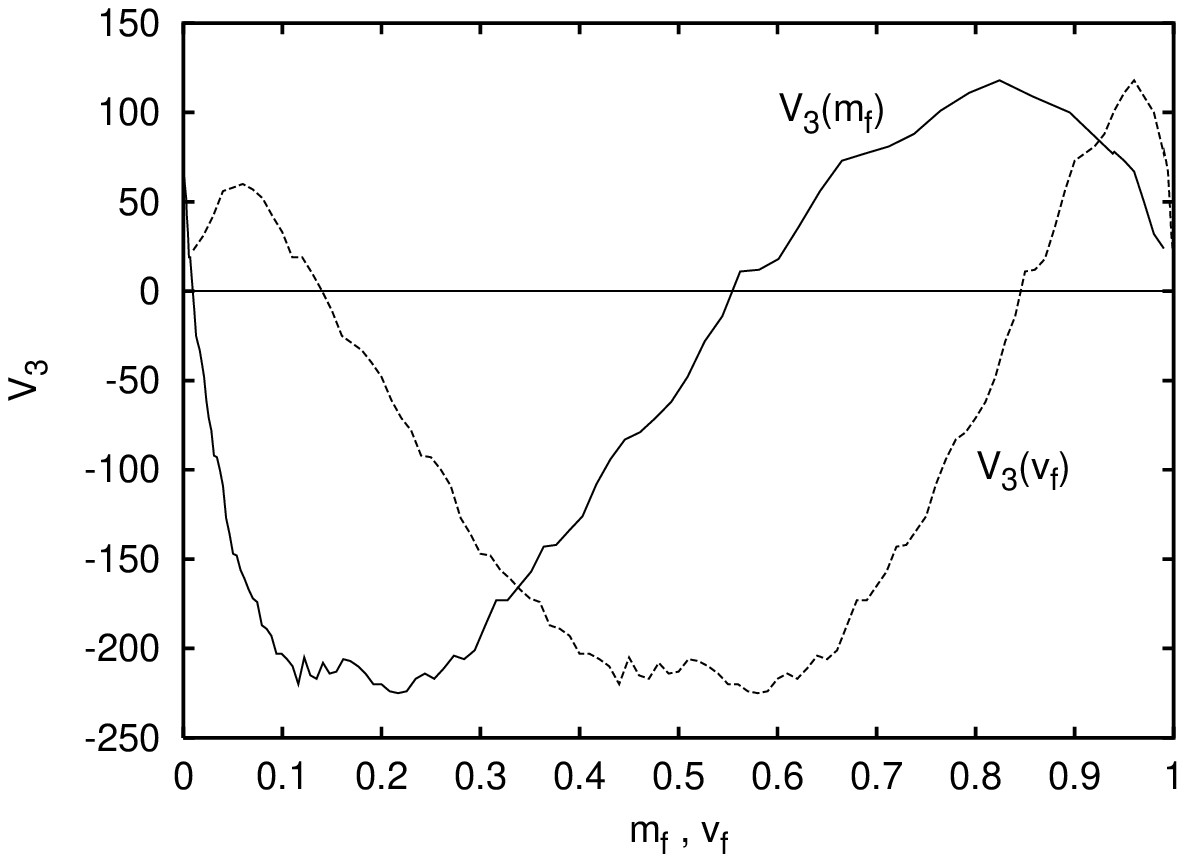}}
  \resizebox{.45\textwidth}{!}{\includegraphics*{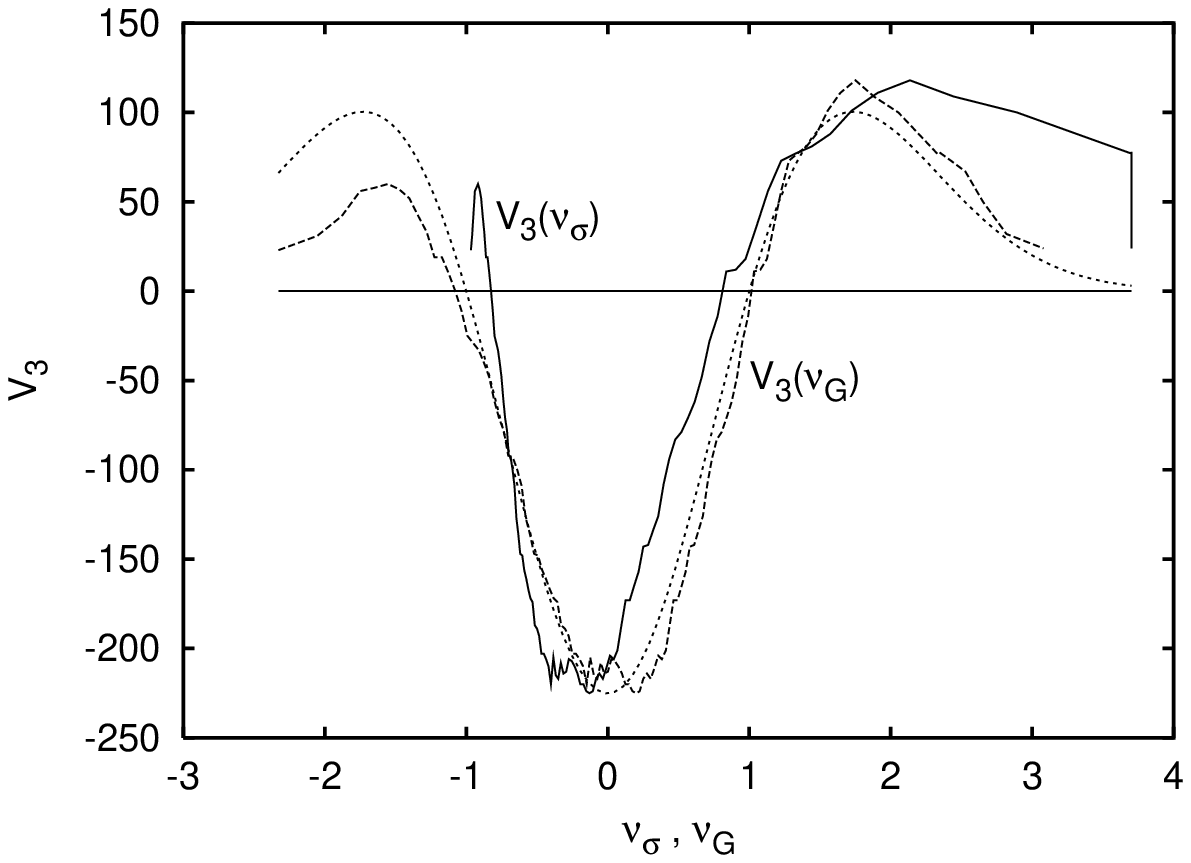}}\\
\caption{The last two Minkowski functionals for N-body model galaxies.
Here $v_f$ is the volume fraction, $m_f$ -- the mass fraction, 
$\nu_\sigma$ -- the normalized volume fraction, and $\nu_G\equiv\nu$ --
the Gaussianized volume fraction from (\ref{nu}).
The dotted line in the right panels
shows the predicted Minkowski functionals for a Gaussian random field MF
\label{fig:MF2_GIF}}
\end{figure}

I show different Minkowski functionals versus different arguments
in Figs.~\ref{fig:MF1_GIF} and \ref{fig:MF2_GIF}. They are calculated
for the model galaxy density distribution shown in the previous figure
(Fig.~\ref{fig:galcube}). Note how much the shape of the same function(al)s  
depends on the arguments used.

\subsection{Gaussian Densities}

All the Minkowski functionals have analytic expressions for
iso-density slices of realizations of Gaussian random fields.
\index{morphology!Gaussian}
For three-dimensional space they are:
\begin{eqnarray}
\label{gaussv}
v_0&=&\frac12-\frac12\Phi\left(\frac{\nu}{\sqrt2}\right)\;,\\
v_1&=&\frac23\frac{\lambda}{\sqrt{2\pi}}\exp\left(-\frac{\nu^2}2\right)\;,\\
v_2&=&\frac23\frac{\lambda^2}{\sqrt{2\pi}}\nu\exp\left(-\frac{\nu^2}2\right)\;,\\
v_3&=&\frac{\lambda^3}{\sqrt{2\pi}}(\nu^2-1)\exp\left(-\frac{\nu^2}2\right)\;,
\end{eqnarray}
where $\Phi(\cdot)$ is the Gaussian error integral, and $\lambda$
is determined by the correlation function $\xi(r)$ of the field:
\beq
\label{lambda}
\lambda^2=\frac1{2\pi}\frac{\displaystyle\xi''(0)}{\displaystyle\xi(0)}\;.
\eeq
The dimension of $\lambda$ is inverse length.

This expression allows to predict all the Minkowski functionals for
a known correlation function (or power spectrum). We can also take a more
empirical approach and to determine $\lambda^2$ on the basis of the
observed density field itself, using the relations
$\xi(0)=\langle\varrho^2\rangle$ and $\xi''(0)=\langle\varrho,_i^2\rangle$,
where $\varrho,_i$ is the density derivative in one coordinate direction.

\begin{figure}
\centering
  \resizebox{.45\textwidth}{!}{\includegraphics*{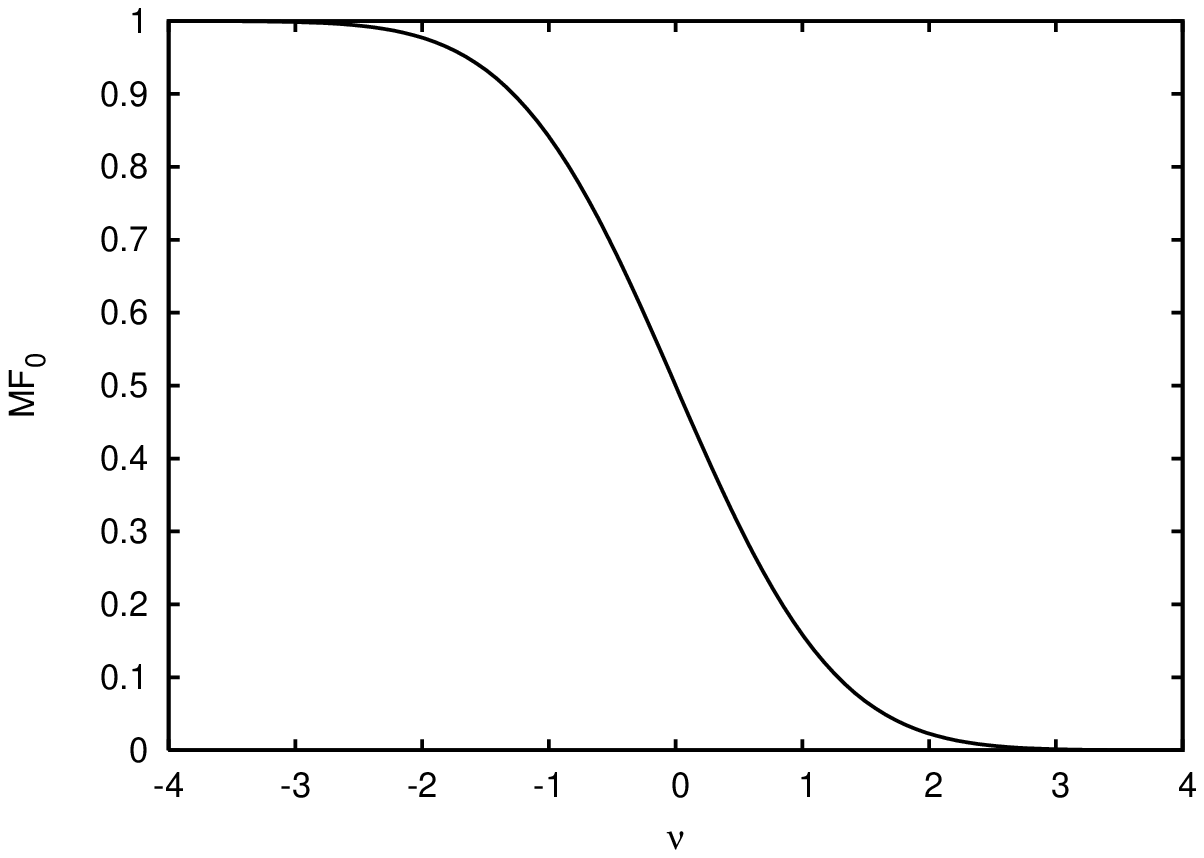}}
  \resizebox{.45\textwidth}{!}{\includegraphics*{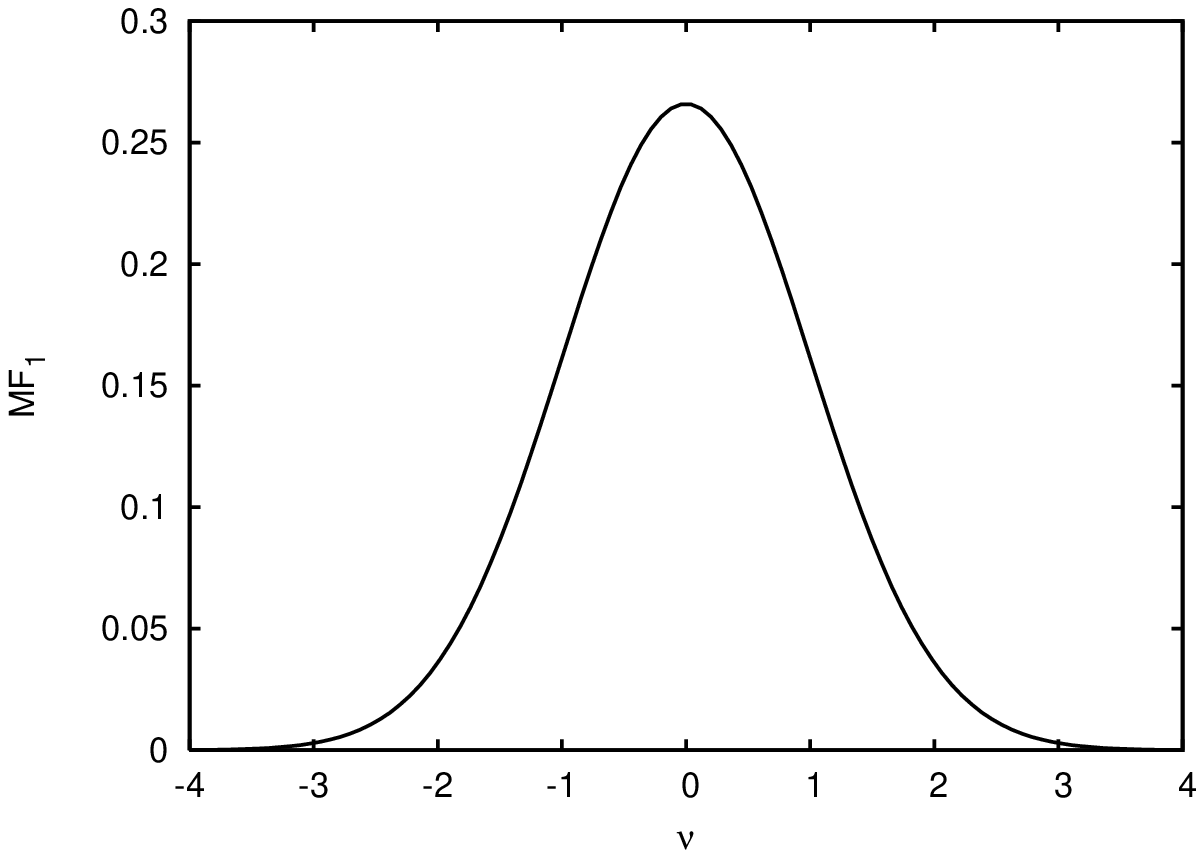}}\\
  \resizebox{.45\textwidth}{!}{\includegraphics*{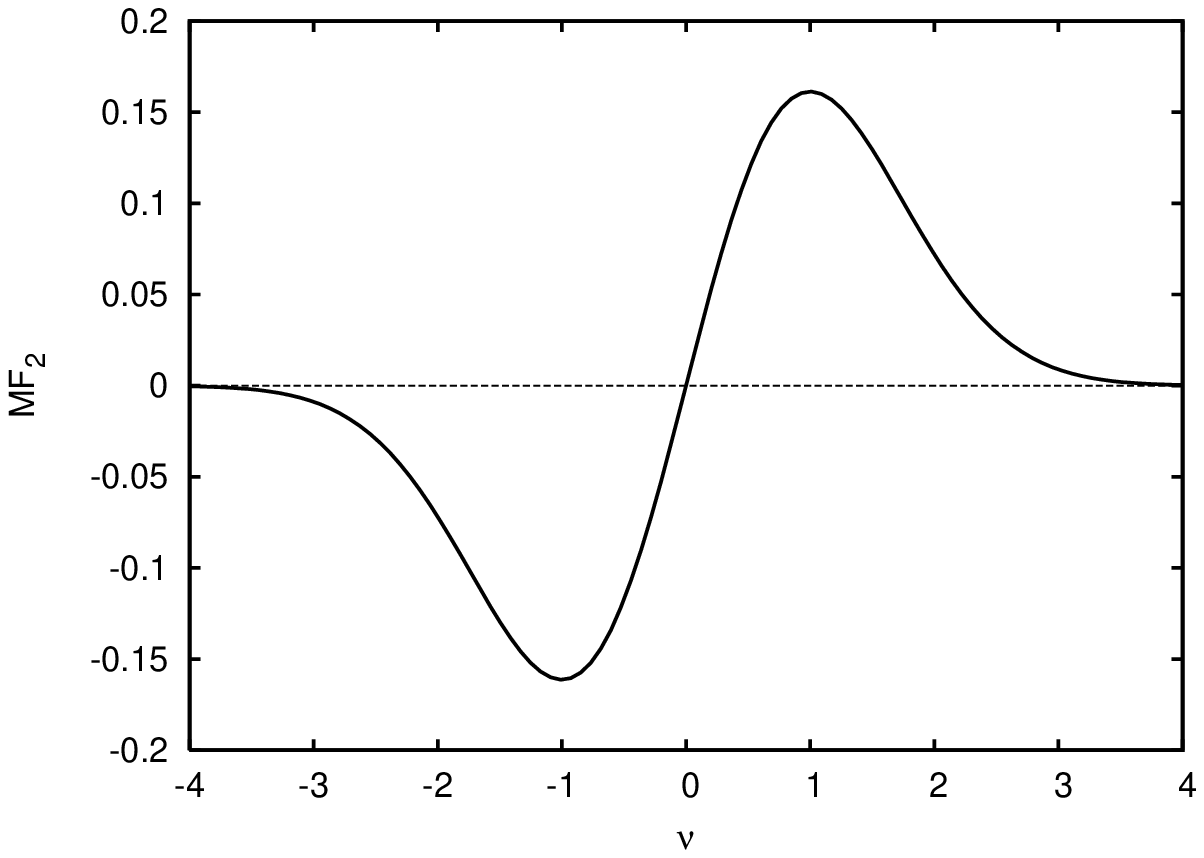}}
  \resizebox{.45\textwidth}{!}{\includegraphics*{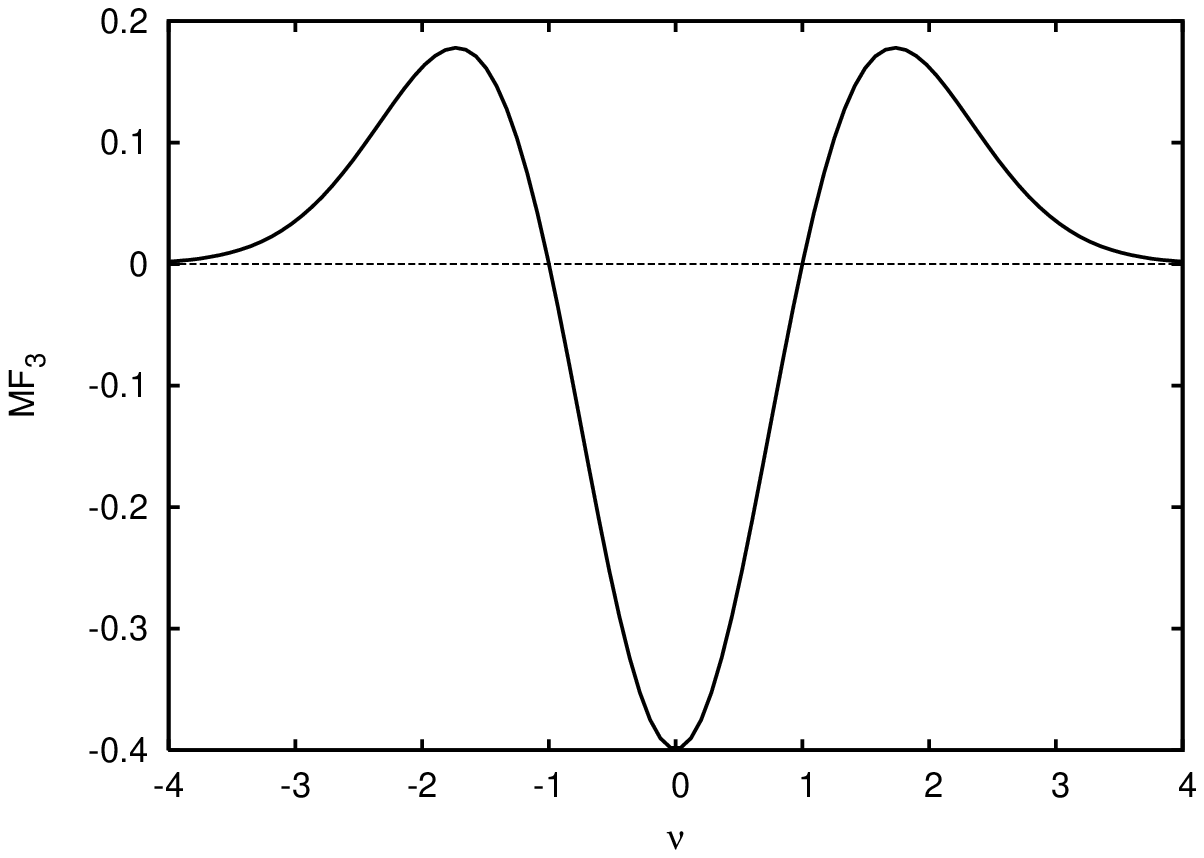}}\\
\caption{Gaussian predictions for Minkowski functionals
$\lambda=1$ \label{fig:MF_gauss}}
\end{figure}

The expected form of these functionals is shown in Fig.~\ref{fig:MF_gauss}.

In practice, it is easy to
obtain good estimates of the Minkowski
functionals for periodic fields. The real data, however, is always
spatially limited, and the limiting surfaces cut the iso-density
surface. An extremely valuable property of Minkowski functionals
is that such cuts can be corrected for -- the data volume mask is another
body, and Minkowski functionals of intersecting bodies can be calculated.
Moreover, if we can assume homogeneity and isotropy for the pattern, we
can correct for border effects of large surveys. This is too technical
for the lecture, so I refer to our forthcoming paper \cite{ESmfpaper2}. 

\subsection{Numerical Algorithms}

Several algorithms are used to calculate the Minkowski functionals
for a given density field and a given density threshold.  We can
either try to follow exactly the geometry of the iso-density surface,
e.g., using triangulation, or to approximate the
excursion set on a simple cubic grid. The first algorithms were
designed to calculate the genus only; the algorithm that was
proposed by Gott, Dickinson, and Melott \cite{ESgott86} uses 
a decomposition of the field
into filled and empty cells, and another class of algorithms 
(Coles, Davies, and Pearson \cite{EScoles96}) uses a grid-valued 
density distribution. The
grid-based algorithms are simpler and faster, but are believed
to be not as accurate as the triangulation codes. 

        I advocate a simple grid-based algorithm, based
\index{algorithms!Crofton}
on integral geometry (Crofton's intersection formula), proposed
by Schmalzing and Buchert (\cite{ESjens}). It  is similar to that of
\cite{EScoles96}, and for $V_3$ (genus) coincides with it.

To start with, we find the density thresholds for given filling
fractions $f$ by sorting the grid densities. This is a common step
for all grid-based algorithms. Vertexes with higher
densities than the threshold form the excursion set. This set is
characterized by its basic sets of different dimensions -- points
(vertexes), edges formed by two neighbouring points, squares (faces)
formed by four edges, and cubes formed by six faces. The algorithm
counts the numbers of elements of all basic sets, and finds the values of the
Minkowski functionals as
\begin{eqnarray}
\label{crofton}
V_0&=&a^3N_3\;,\nonumber\\
V_1&=&a^2\left(\frac29N_2-\frac23N_3\right)\;,\nonumber\\
V_2&=&a\left(\frac29N_1-\frac49N_2+\frac23N_3\right)\;,\nonumber\\
V_3&=&N_0-N_1+N_2-N_3\;,
\end{eqnarray}
where $a$ is the grid step, $N_0$ is the
number of vertexes, $N_1$ is the number of edges, $N_2$ is the
number of squares (faces), and $N_3$ is the number of basic cubes in
the excursion set.

This algorithm described above is simple to program, and is very fast,
allowing the use of Monte-Carlo simulations for error estimation
\footnote{A thorough analysis of the algorithm and its application
to galaxy distributions will be available, by the time this book will
be published (\cite{ESmfpaper2}).}.

The first, and most used, algorithm was '\texttt{CONTOUR}' and its
derivations; this algorithm was written by David Weinberg and was probably
one of the first cosmological public-domain algorithms \cite{EScontour}.
However, there is a noticeable difference between the results 
that our grid algorithm produces and 
those of '\texttt{CONTOUR}'. index{algorithm:CONTOUR}
You can see it yourself, 
comparing the Minkowski functional
figures in this lecture with Fig.~4 in Peter Coles' chapter. All genus
curves ever found by '\texttt{CONTOUR}' look like that there, 
very jaggy. How much I
have tried, using Gaussian smoothing and changing smoothing lengths, I have
never been able to reproduce these jaggies by our algorithm. 
At the same time, the
calculated genus curves always follow the Gaussian prediction, so there
is no bias in either algorithm. As the genus ($V_3$) counts
objects (balls, holes, handles), it should, in principle, have mainly 
unit jumps, and only occasionally a larger jump. That is what we see in
our genus ($V_3$) graphs. The only conclusion
that I can derive is that the algorithm we use is more stable,
with a much smaller variance.

\subsection{Shapefinders}

\index{shapefinders}
As Minkowski functionals give a complete morphological
description of surfaces, it should be possible to use them
to numerically describe the shapes of objects, right? For
example, to differentiate between fat (spherical) objects
and thin (cylindrical objects), banana-like superclusters and
spiky superclusters. This hope has never died, and different shape
descriptors have been proposed (a selection of them is listed in
our book \cite{ESmartsaar}). The set of shape
descriptors that use Minkowski functionals was proposed by
Sahni, Sathyaprakash and Shandarin \cite{ESsahni}, and is called 'shapefinders'.
Now, shapefinder definitions have a habit of changing
from paper to paper and it is not easy to
follow the changes, so I shall give here a careful derivation
of the last version of shapefinders. 
Shapefinders are defined as ratios of numbers that are similar to
Minkowski functionals, but only close. in 3-D, the chain of Minkowski
functionals $V_i$, from $V_0$ to $V_3$, has gradually diminishing
dimensions, from $L^3$ to $L^0$. So, the ratios of neighbours in this
chain have a dimension of length; these ratios make up the first set
of shapefinders. The proportionality coefficients are chosen to
normalize all these ratios to $R$ for a sphere of the radius $R$.
Take a surface that is delimiting (shaping) a three-dimensional
object (e.g., a supercluster).
The definitions of the first three Minkowski functionals (\ref{mf0}--\ref{mf2})
can be rewritten as
\begin{eqnarray*}
V&=&V_0=\frac{4\pi}3R^3\;,\\
S&=&6V_1=4\pi R^2\;,\\
C&=&3\pi V_2=R\;,\\
\end{eqnarray*}
where the second equalities stand for a sphere of a radius
$R$, $V$ is the volume, $S$ -- the surface, and $C$ -- the
mean integrated radius of curvature:
\beq
\label{C}
C=\int_S \frac12\left(\frac1{R_1(\mathbf{x})}+\frac1{R_2(\mathbf{x})}\right)
		\,dS\;.
\eeq
Whatever formulae you see in the literature, do not use them
before checking with the definition below -- these are the
true shapefinders:
\begin{eqnarray}
\label{shape}
H_1&=&\frac{3V}{S}=\frac12\frac{V_0}{V_1}\;,\quad\mathrm{thickness}\;,\\
H_2&=&\frac{S}{C}=\frac2{\pi}\frac{V_1}{V_2}\;,\quad\mathrm{breadth}\;,\\
H_3&=&\frac{C}{4\pi}=\frac34 V_2\;,\quad\mathrm{length}\;.\\
\end{eqnarray}
\index{shapefinders!thickness} \index{shapefinders!breadth} 
\index{shapefinders!length}
Only this normalization gives you $H_i=R$ for a sphere; check it.
The descriptive names you see were given by the authors.
There is a fourth shapefinder -- the genus has been given the honour to
stand for it, directly. As genus counts ``minus'' things (minus the number
of isolated objects, minus the number of holes), the fourth Minkowski
functional $V_3$ should be a better candidate.

The first set of shapefinders is accompanied by the 'second order'
shapefinders
\begin{eqnarray}
\label{K}
K_1&=&\frac{H_2-H_1}{H_2+H_1}\;,\quad\mathrm{planarity}\;,\\
K_2&=&\frac{H_2-H_1}{H_2+H_1}\;,\quad\mathrm{filamentarity}\;.\\
\end{eqnarray}
\index{shapefinders!planarity}
\index{shapefinders!filamentarity}
These five (six, if you count the genus) numbers describe pretty well
the shape of smooth (ellipsoidal) surfaces.  In this case, the
ratios $K_1$ and $K_2$ vary nicely from 0 to 1, and another
frequently used ratio, $K_1/K_2$, has definite trends with respect to the
shape of the object. But things get much more interesting when you start
calculating the shapefinders of real superclusters; as shapes get
complex, simple meanings disappear. Also, as shapefinders are defined as
ratios of observationally estimated numbers (or ratios of quadratic
combinations, as $H_i$ work out in terms of $V_i$), they are
extremely noisy. So, in case of serious use, a procedure should be
developed to estimate the shapefinders directly, not by the ratio rules
(\ref{shape}, \ref{K}); such a procedure does not exist yet.
But, for the moment, the shapefinders
are probably the best shape description tool (for cosmology) we have.

\subsection {Morphology of Wavelet-Denoised Density}

\index{density!wavelet-denoised}
So much for the preliminaries. Now we have all our tools 
(wavelets, densities, Minkowski functionals). Let us use them and see
what is the morphology of the real galaxy density distribution.

This question has been asked and answered about a hundred times,
starting from the first topology paper by Gott et al \cite{ESgott86}.
The first data set was a cube from the CfA I sample and contained
123 galaxies, if I remember right. (Imagine estimating a spatial
density on the basis of a hundred points; this is the moment that
Landau's definition of a cosmologist is appropriate: ``Cosmologists
are often wrong, but never in doubt''.) The answer was -- Gaussian\,!\,;
ten points for courage.
The same  optimistic answer has been heard about a hundred times
since (in all the papers published), with slight corrections in
later times, as data is getting better. These corrections have been explained
by different observational and dynamical effects, and peace reigns. 

But -- all these studies have carefully smoothed the galaxy catalogues
by nice wide Gaussian kernels to get a proper density. Doubts have been
expressed that one Gaussianity could lead to another (by Peter Coles,
for example), and the nice Gaussian behaviour we get is exactly that
we have built in in the density field. So, let us wavelet-denoise the
density field (a complex recipe, but without any Gaussians), and
estimate the Minkowski functionals. The next two figures are from
our recent work (\cite{ESmfpaper1}).

\begin{figure}
\centering
\resizebox{.8\textwidth}{!}{\includegraphics*{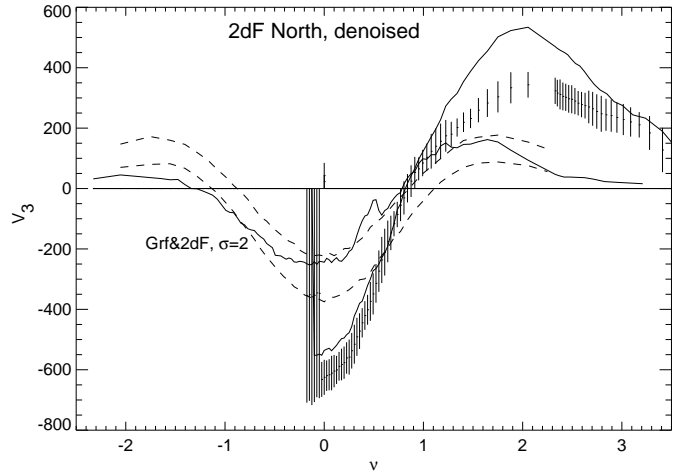}}
  \caption{The Minkowski functional $V_3$ for the 2dF brick, for
the wavelet denoised data
set (thick solid line). The variability range of the
wavelet denoised mocks is shown with bars. We show also the 95\% confidence 
limits
for 1300 realizations of theoretical Gaussian density fields (dashed lines),
and the $V_3$ data curve (thin solid line), all obtained for the 
Gaussian $\sigma=2$\,Mpc/$h$ smoothing
\label{fig:mfclean}}
\end{figure}

The data are the 2dFGRS Northern galaxies; we \index{sample!2dFGRS North}
chose a maximum-volume one-magnitude interval volume-limited
\index{sample!volume-limited}
(constant density) sample and cut a maximum-volume brick from it
(interesting, every time I say 'maximum', the sample gets smaller).
Constant density is necessary, otherwise typical density scales
will change with distance, and a
brick was cut in order to avoid border corrections
(these can bring additional difficulties for wavelet denoising). 
We carefully wavelet-denoised the density; Fig.~\ref{fig:mfclean}
shows its fourth Minkowski functional.

\begin{figure}
\centering
\resizebox{.8\textwidth}{!}{\includegraphics*{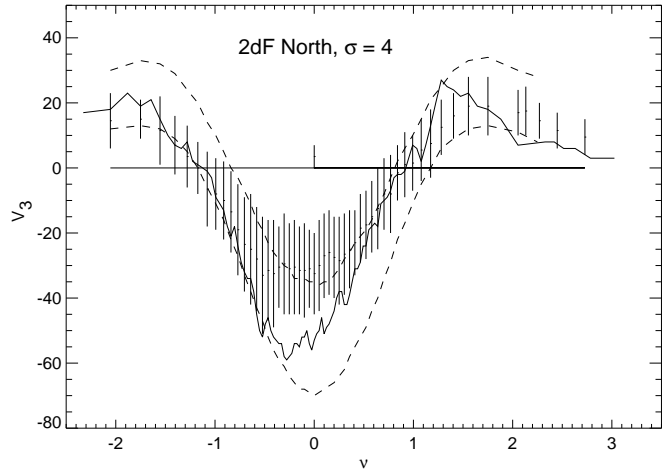}}
  \caption{The Minkowski functional $V_3$ for the 2dF brick, 
Gaussian smoothed with $\sigma=4$\,Mpc/$h$
(thick solid line). The variability range of the
mocks is shown with bars. We show also the 95\% confidence 
limits
for 1300 realizations of theoretical Gaussian density fields (dashed lines).
\label{fig:gauss4}}
\end{figure}

First, we see that our wavelet.denoised density is never close to Gaussian.
Secondly, the specially built N-body catalogues (mocks) do not have
Gaussian morphology, too; although they deviate from the real data,
they are closer to it than to the example Gaussian. We see also that the
same galaxies, smoothed by a yet very narrow Gaussian kernel
($\sigma=2$\,Mpc/$h$), show an
almost Gaussian morphology. Thus, the clear message of this figure is that
the morphology of a good (we hope the best) adaptive galaxy density
is far from Gaussian; and the Gaussianity-confirming results obtained so
far are all the consequence of the Gaussianity input by hand -- Gaussian
smoothing. This figure exhibits a little non-Gaussianity yet; the next one 
(Fig.~\ref{fig:gauss4}) almost does not. The filter used there is still
narrow ($\sigma=4$\,Mpc/$h$); the filter widths used in most papers
are around twice that value.

I am pretty sure that Gaussian smoothing is the culprit here. We built
completely non-Gaussian density distributions (Voronoi walls and networks),
Gaussian smoothed them using popular recipes about the kernel size, 
and obtained perfectly Gaussian Minkowski functionals. 

One reason for that, as Vicent Mart{\'\i}nez has proposed, is that
the severe smoothing used changes a density field into Poissonian,
practically. Try it -- smooth a density field with a Gaussian of $\sigma=r_0$,
where $r_0$ is the correlation length, and you get a density field with
a very flat low-amplitude correlation function, almost Poissonian. 
And the Minkowski functionals of Poissonian density 
\index{morphology!Poissonian}
fields\footnote{To be more exact,
here is the recipe -- take a Poissonian point process of $N$ points in a
volume $V$ and smooth it with a Gaussian kernel of $\sigma=d$, where 
$d=(V/N)^{1/3}$ is the mean distance between the points.} are Gaussian;
we tested that.

Another reason that turns Minkowski functionals Gaussian even for small
$\sigma$, must be the extended wings of Gaussian kernels. Although they drop
pretty fast, they are big enough to add to a small extra ripple on the
main density field. As Minkowski functionals are extremely sensitive
to small density variations, then all they see is that ripple.
This is especially well seen in initially empty regions. Usually,
one uses a FFT-based procedure for Gaussian smoothing, as that is
at least a hundred times faster (for present catalogue volumes) that
direct convolution in real space. This procedure generates a wildly
fluctuating small-amplitude density field in empty regions, and we
did not realize for a long time at first where those giant-amplitude
ghost MF-s came from. Gaussian smoothing and FFT, that was their address.

\subsection {Multiscale Morphology}

\index{multiscale!morphology} \index{morphology!multiscale}
So, the galaxy density, similar to that  we have accustomed 
to find in our everyday experience, is decidedly non-Gaussian.
But is that a problem? Cosmological dynamics tells us that structure
evolves at different rate at different scales; a true density mixes
these scales all together and is not the best object to search for 
elusive traces of initial conditions. Good. Multiscale densities to
the rescue.

	The results that will end this chapter did not exist at the
time of the summer school. But we live in the present, and time
is short, so I will include them. A detailed account is already accepted
for publication (\cite{ESmfpaper2}), I shall show only a collection of
Minkowski functionals here.

As our basic data set, we took the 2dFGRS volume-limited samples
for the $[-20,-19]$ magnitude interval; they have the highest mean
density among similar one-magnitude interval samples. We did not
cut bricks this time, but corrected for sample boundaries; we have
learnt that by now. We wavelet-decomposed the galaxy density fields
and found the Minkowski functionals; as simple as that. Although
wavelet decompositions are linear and should not add anything to
the morphology of the fields, we checked that on simulated Gaussian
density fields. Right, they do not add any extra morphological signal.

\begin{figure}
\centering
\resizebox{.8\textwidth}{!}{\includegraphics*{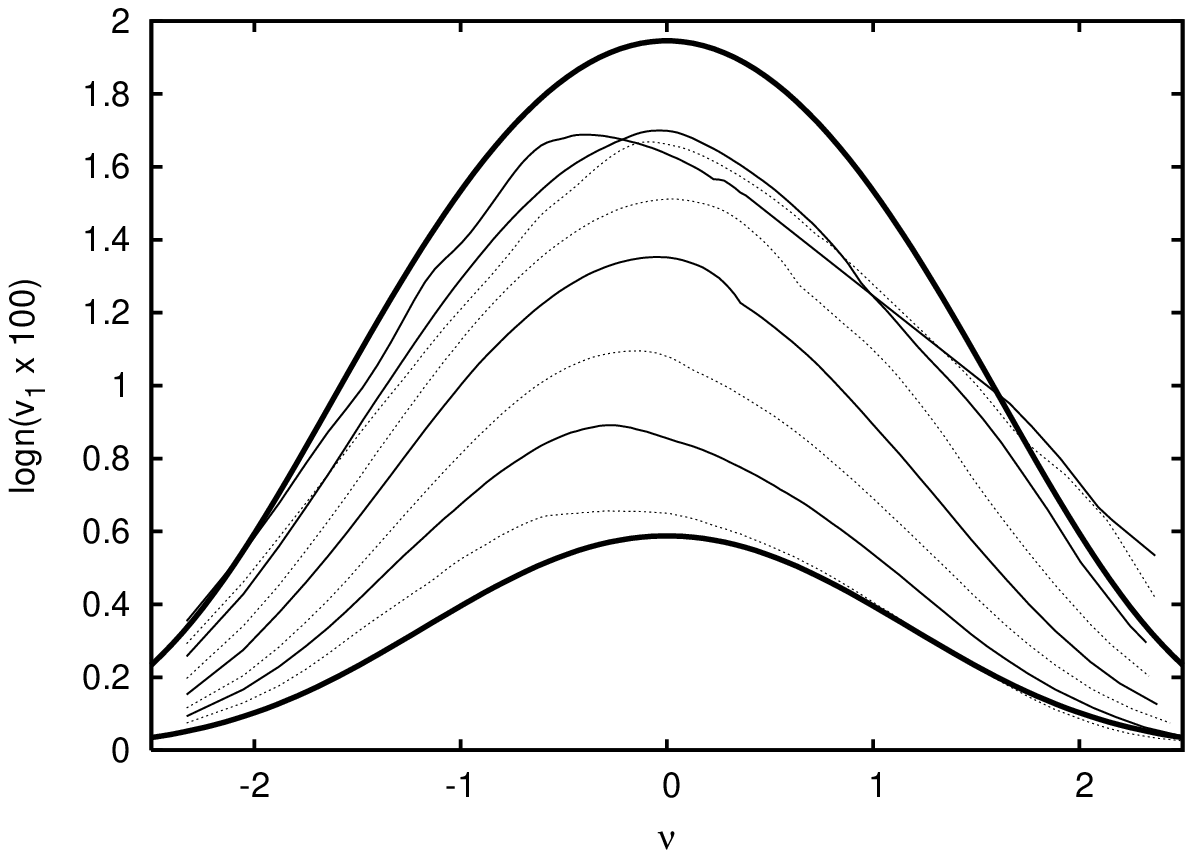}}
\caption{Summary of the densities of the second MF $v_1$ for 
the data and all wavelet orders for the 2dFN19 sample, in the
logn mapping. Thick lines show reference Gaussian predictions.
Full lines stand for the 1\,Mpc/$h$ grid, dotted lines -- for the
$\sqrt2$\,Mpc/$h$ grid
\label{fig:2dfNmf1}}
\end{figure}

\begin{figure}
\centering
\resizebox{.8\textwidth}{!}{\includegraphics*{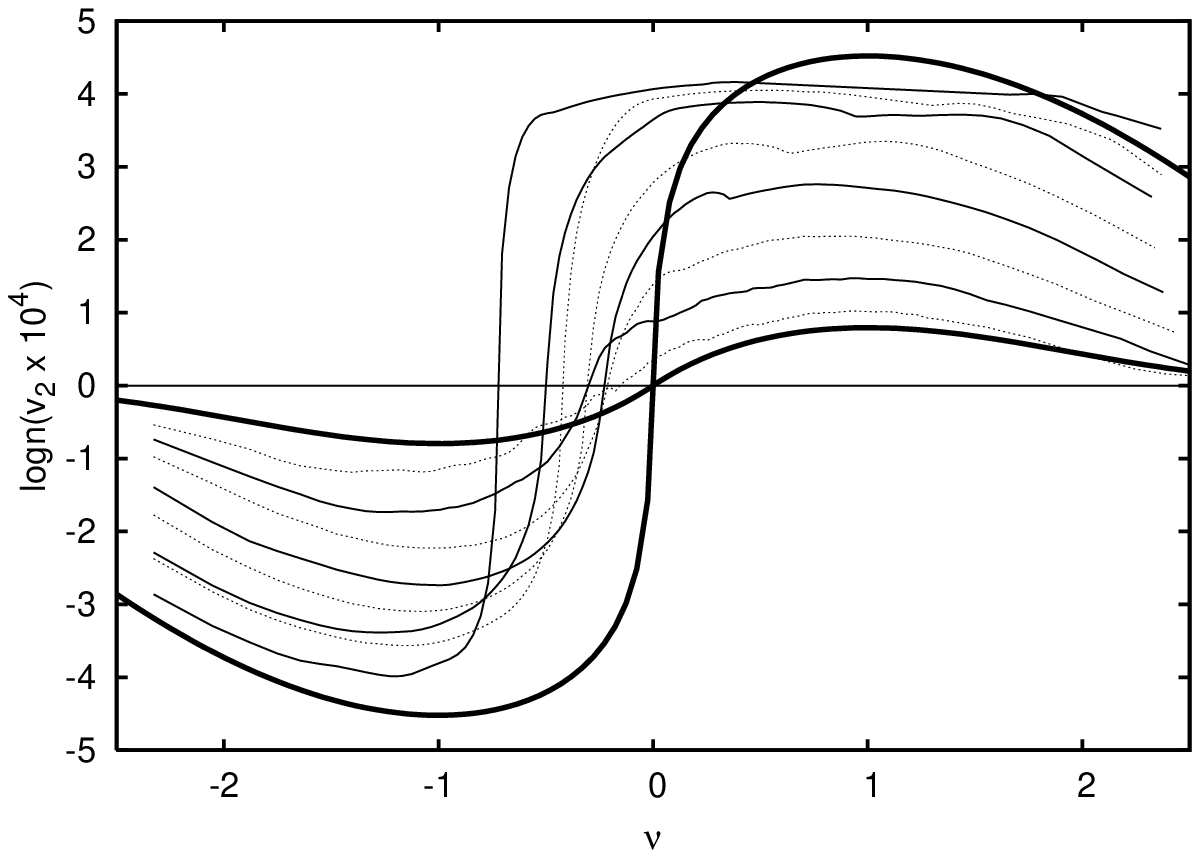}}
\caption{Summary of the densities of the third MF $v_2$ for 
the data and all wavelet orders for the 2dFN19 sample, in the
logn mapping. Thick lines show reference Gaussian predictions.
Full lines stand for the 1\,Mpc/$h$ grid, dotted lines -- for the
$\sqrt2$\,Mpc/$h$ grid
\label{fig:2dfNmf2}}
\end{figure}

\begin{figure}
\centering
\resizebox{.8\textwidth}{!}{\includegraphics*{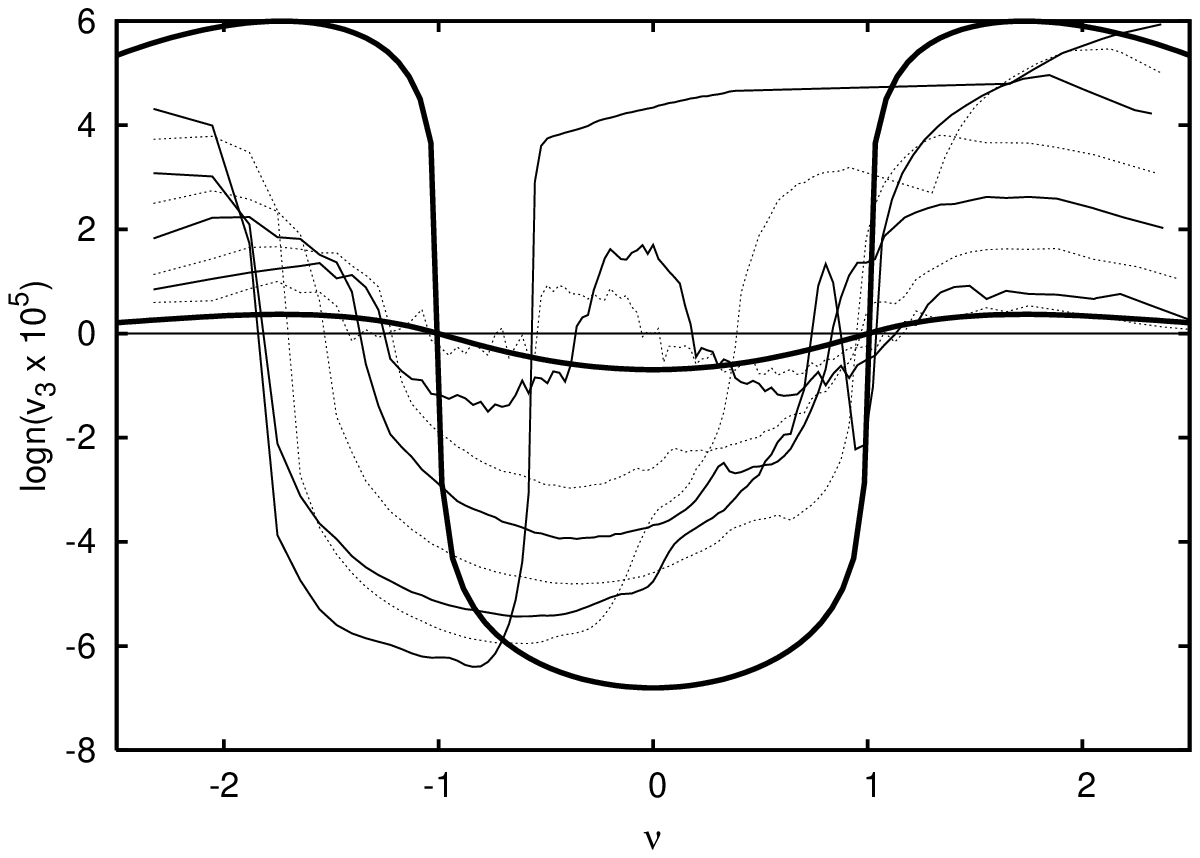}}
\caption{Summary of the densities of the fourth MF $v_3$ for 
the data and all wavelet orders for the 2dFN19 sample, in the
logn mapping. Thick lines show reference Gaussian predictions.
Full lines stand for the 1\,Mpc/$h$ grid, dotted lines -- for the
$\sqrt2$\,Mpc/$h$ grid
\label{fig:2dfNmf3}}
\end{figure}

The results (for the 2dFGRS North) are shown in 
Figs.~\ref{fig:2dfNmf1}--\ref{fig:2dfNmf3}. As the amplitudes of
the (densities of) Minkowski functionals vary in a large interval,
we use a sign-aware logarithmic mapping:
\[
\mathrm{logn}(x;a)=\mathrm{sgn}(x)\log(1+|x|/a)\;.
\]
This mapping accepts arguments of both signs ($\log(x)$ does not),
is linear for $|x|<<a$ and logarithmic for $|x|>>a$.
As the figures show, for the scales possible to study, the morphology
of the galaxy density distribution is decidedly not Gaussian.
The deviations are not too large for the second Minkowski functional $v_2$
(watch how the maxima shift around),
but are clearly seen for the two others. The maximum wavelet order here is
3 for a $\sqrt2$\,Mpc/$h$ grid, that corresponds to a characteristic scale
of $2^3\sqrt2\approx 11.3$\,Mpc/$h$. As the mean thickness of the
2dFGR North slice is about 40--50\,Mpc/$h$, we cannot go much further --
the higher order wavelet slices would be practically two-dimensional.
The 2dFGRS Southern dataset has similar size limitations. So, as
10\,Mpc/$h$ is a scale where cosmological dynamics might have slight
morphological effects, the question if the original morphology of
the cosmological density field was Gaussian, is not answered yet. But
we shall find it out soon, when the SDSS will finally fill its full 
planned volume. 

These are the results, for the moment. The morphology of the galaxy
density field is far from Gaussian, in contrary to practically every
earlier result. Have all these results only confirmed that if we take
a serious smoothing effort, we are able to smooth any density field
to Poissonian? An indecent thought.

But do not be worried, 
results come and go, this is the nature of research. Methods, on the
contrary, stay a little longer, and this school was all about methods.

\section*{Recommended Reading}

I was surprised that Bernard Jones recommended a wavelet bookshelf
that is almost completely different from mine (the only common book
is that of I.~Daubechies'). So, read those of
Bernard's choice, and add mine:
\begin{itemize}
\item five books:
  \begin{enumerate}
  \item  Stephane Mallat, A Wavelet Tour of Signal Processing, 2nd ed.,
  Academic Press, London,  1999,
  \item C. Sidney Burrus, Ramesh A. Gopinath, Haitao Gao, Introduction
  to Wavelets and Wavelet Transforms, Prentice Hall, NJ, 1997,
  \item Ingrid Daubechies, Ten lectures on wavelets, SIAM, Philadelphia,
  2002, 
  \item Jean-Luc Starck, Fionn Murtagh, ``Astronomical Image
  and Data Analysis'', 2nd ed., Springer, 2006 (application of wavelets 
  and many other wonderful image processing methods in astronomy),
  \item Bernard W. Silverman, Density Estimation for Statistics and
  Data Analysis, Chapman \& Hall / CRC Press, Boca Raton, 1986 (the classical
  text on density estimation),
  \end{enumerate}
\item three articles:
  \begin{enumerate}
  \item Mark J. Shensa, The Discrete wavelet Transform: Wedding the \`A Trous 
  and Mallat Algorithms, IEEE Transactions on Signal Processing, 40, 2464--2482,
  1992 (the title explains it all),
  \item K.R. Mecke, T. Buchert, H. Wagner, Robust morphological measures for 
  large-scale structure in the Universe, Astron. Astrophys. 288, 697--704,
  1994 (introducing Minkowski functionals in cosmology),
  \item Jens Schmalzing, Thomas Buchert, Beyond Genus Statistics: 
  A Unified Approach
  to the Morphology of Cosmic Structure, Ap. J. Letts  482, L1, 1997,
  (presentation of two grid algorithms).
  \end{enumerate}
\item two web pages:
  \begin{enumerate}
  \item a wavelet tutorial by Jean-Luc Starck at 
  \url{(http://jstarck.free.fr)}, 
  \item the wavelet pages by David Donoho at 
  \url{http://www-stat.stanford.edu/~donoho/} (look at
  lectures and reports).
  \end{enumerate}
\end{itemize}

\section*{Acknowledgements}
I was introduced to wavelets in about 1990 by Ivar Suisalu 
(he was my PhD student then).
As that happened in NORDITA, I asked advice from Bernard Jones soon,
and started on a wavelet road, together with Vicent Mart{\'\i}nez 
and Silvestre Paredes; these early wavelets were continuous. 
Much later, I have returned to wavelets, and have learnt much from 
Bernard, who is using wavelets in the real world, and from
Vicent, Jean-Luc Starck, and David Donoho, the members of
our multiscale morphology group.
I thank them all for pleasant collaboration and knowledge shared.
All the results presented here belong to our morphology group.

My present favourites are the \emph{\`a trous} wavelets, as you have noticed.

My research has been supported in Estonia by the Estonian Science Foundation 
grant 6104, and by the Estonian Ministry of Education research project 
TO-0060058S98. In Spain, I have been supported by the University of Valencia
via a visiting professorship, and by the Spanish MCyT project 
AYA2003-08739-C02-01 (including FEDER).

\end{document}